\documentclass[lettersize,journal]{IEEEtran}
\usepackage{amsmath,amssymb,amsfonts}
\usepackage{textcomp}
\usepackage{xcolor}

\usepackage{algorithmic}
\usepackage{array}
\usepackage[caption=false,font=normalsize,labelfont=sf,textfont=sf]{subfig}
\usepackage{textcomp}
\usepackage{stfloats}
\usepackage{verbatim}
\usepackage{graphicx}
\usepackage{cite}
\usepackage{amsthm}
\usepackage{array, multirow}
\usepackage{pifont}

\usepackage[linesnumbered,procnumbered,ruled,vlined]{algorithm2e}
\usepackage{kotex} 
\usepackage{stfloats}
\usepackage{url}

\usepackage{verbatim}\usepackage{xspace}
\usepackage[autostyle]{csquotes}
\usepackage{float}
\usepackage[flushleft]{threeparttable} 
\usepackage{epsfig}
\usepackage{booktabs}

\newtheorem{lemma}{Lemma} 
\newtheorem{theorem}{Theorem}[section]
\newtheorem{definition}{Definition}[section]
\newtheorem{assumption}{Assumption}[section]
\newtheorem{corollary}{Corollary}[section]

\begin{document}

\title{A Hierarchical Sharded Blockchain Balancing Performance and Availability}

\author{Yongrae Jo and Chanik Park
    \thanks{Manuscript received April 19, 2021; revised August 16, 2021.}
    \thanks{This work was supported by Institute of Information \& Communications Technology Planning \& Evaluation (IITP) grant funded by the Korea government(MSIT) (No. 2021-0-00484, Core Technologies for Balanced P2P Network-based Blockchain Services; No. 2020-0-00936, Core Technologies for 5G-Aware Blockchain Networks; No. 2021-0-00136, Development of Big Blockchain Data Highly Scalable Distributed Storage Technology for Increased Applications in Various Industries)}
    \thanks{Yongrae Jo and Chanik Park are with Pohang University of Science and Technology, Pohang, Republic of Korea. (e-mail: memex@postech.ac.kr; cipark@postech.ac.kr)}
}

\markboth{Journal of \LaTeX\ Class Files,~Vol.~14, No.~8, August~2021}%
{Shell \MakeLowercase{\textit{et al.}}: A Sample Article Using IEEEtran.cls for IEEE Journals}

\IEEEpubid{0000--0000/00\$00.00~\copyright~2021 IEEE}

\maketitle

\begin{abstract}
    Blockchain networks offer decentralization, transparency, and immutability for managing critical data but encounter scalability problems as the number of network members and transaction issuers grows.
    While sharding enhances blockchain scalability, most existing techniques prioritize performance over availability(e.g., a failure in a few servers holding a shard leads to data unavailability).
    In this paper, we propose PyloChain, a hierarchical sharded blockchain that balances availability and performance.
    PyloChain consists of multiple lower-level local chains and one higher-level main chain. Each local chain speculatively executes local transactions to achieve high parallelism across multiple local chains.
    The main chain leverages a directed-acyclic-graph (DAG)-based mempool to guarantee local block availability and to enable efficient Byzantine Fault Tolerance (BFT) consensus to execute global (or cross-shard) transactions within collocated shards.
    In order to reduce the number of aborted local transactions, PyloChain applies a simple, but effective scheduling technique to handle global transactions in the main chain.
    PyloChain provides a fine-grained auditing mechanism to mitigate faulty higher-level members by externalizing main chain operations to lower-level local members.
    We implemented and evaluated PyloChain, demonstrating improved overall performance scalability. For example, under 20\% global transactions and 12 sharding zones, PyloChain achieves 1.49x higher throughput and 2.63x lower latency compared to the balanced sharding baseline.
\end{abstract}

\begin{IEEEkeywords}
    Blockchain, Sharding, Availability, Performance
\end{IEEEkeywords}

\section{Introduction}

Blockchain networks support decentralization, transparency, and immutability to manage critical data.
However, as the number of network members and transaction-issuing users increases, it faces a scalability problem in efficiently handling a large volume of transactions.
Sharding-based parallelism \cite{SharPer, Saguaro, Ziziphus, AHL, Monoxide, BrokerChain, kokoris2018omniledger, rapidchain, SharDag,Benzene,HieraChain,FSBlockchain, Hiba,GriDB, Monoxide, Chainspace, Meepo, MeepoJournal, PShard, OHIE, dylochain} offers a promising solution by partitioning the blockchain to process transactions concurrently. Two primary schemes exist: availability-favored and performance-favored sharding. Availability-favored sharding \cite{Meepo,MeepoJournal,PShard,OHIE} replicates a full copy of shards (i.e., collocated) across every server, allowing parallel consensus among corresponding shards, ensuring data availability and efficient cross-shard transaction executions. Performance-favored sharding \cite{SharPer, Saguaro, Ziziphus, AHL, Monoxide, BrokerChain, kokoris2018omniledger, rapidchain, SharDag,Benzene,HieraChain,FSBlockchain} replicates a single copy of a shard (i.e., isolated) to a disjoint group of servers, enhancing performance by reducing storage, networking, and consensus overhead within each server.

\captionsetup[subfloat]{font=scriptsize,labelfont=small} 


However, the availability-favored sharding suffers from limited scalability because it requires all servers to hold a complete copy of the shards. This significantly burdens each member as the number of members and shards increases. Meanwhile, performance-favored sharding may pose risks to data availability because it inherently replicates data to fewer servers, which means that compared to replicating data across all servers, the availability of the entire system data can be severely compromised even if only a few servers fail.
These constraints considerably hinder the practical use of sharded blockchains in real-world business operations.
Thus, it is desirable to motivate the need for a balanced approach where the previous two approaches are combined to find a sweet spot between availability and performance.

A hierarchical blockchain architecture with sharding zones presents an effective way to realize this balanced approach.
By organizing the network into hierarchical sharded zones, each full member, which participates representatively in cross-zone communication, manages the higher-level main chain (i.e., a full copy of all shards). Meanwhile, local members within each zone manage the lower-level local chain (i.e., a single shard). This architecture ensures data availability even if a specific shard fails, while still benefiting from performance advantages of sharding-based parallelism. Additionally, a hierarchical sharded blockchain can efficiently process cross-shard transactions by exploiting collocated shards within full members.

We identify three key challenges in designing the hierarchical blockchain architecture:
First, zone scalability; For data availability, lower-level local blocks need to be reliably propagated via cross-zone communications over an asynchronous network and integrated into the higher-level main chain. As the number of shards increases, large volumes of local blocks from multiple zones are generated concurrently, making the scalability of the main chain consensus crucial.
Second, transaction scalability: As a very large volume of local blocks from many sharding zones is integrated into the main chain, efficiently processing these transactions becomes increasingly important. In particular, global (or cross-shard) transactions may abort local transactions that have been speculatively executed within a local chain, increasing synchronization overhead across the hierarchy.
Third, malicious full member: Each full member plays a critical role in cross-zone communications, responsible for relaying essential data across the hierarchy. The main challenge arises from the fact that local members are unaware of what happens on the main chain, which malicious full members may exploit.
Therefore, auditing the trustworthiness of full members is crucial for correct hierarchical operation.


In this paper, we propose PyloChain, a hierarchical sharded blockchain that balances availability and performance by addressing the key challenges.
For the first challenge, PyloChain employs a DAG-based mempool \cite{narwhaltusk} to append highly concurrent local blocks from multiple sharding zones simultaneously in a leaderless manner, ensuring local block availabilityas well as their efficient total ordering with high throughput.
To the second challenge, PyloChain adopts and enhances the order-execute-order-validate (O-X-O-V) transaction processing model \cite{dylochain} by incorporating a simple scheduling technique. This model enables speculative parallel execution of local transactions across zones and batch-processing of cross-shard transactions at the end of each main block processing cycle to reduce aborted transactions, significantly reducing the burden on main block processing.
For the third challenge, we design a fine-grained auditing mechanism that externalizes the full member's critical actions on the main chain, allowing them to be audited by local members. To achieve this, PyloChain divides the operational semantics of the DAG-based main chain protocol into fine-grained phases, enabling asynchronous audits to verify the full member's behavior at each phase.

We implement and evaluate PyloChain, demonstrating that it achieves practical performance across various workloads in a network with up to 18 zones, including 18 full members and 72 local members. We observed that PyloChain consistently demonstrates better performance scalability compared to the state-of-the-art balanced hierarchical sharded blockchain \cite{dylochain}. Specifically, under 20\% global transactions and 12 zones, PyloChain achieved 1.49x higher throughput and 2.63x faster latency. Also, we show that PyloChain effectively balances performance and availability, by comparing the other sharding schemes.


In summary, our key contributions are:
\begin{itemize}
    \item We propose PyloChain, a hierarchical sharded blockchain balancing performance and availability.
    \item PyloChain leverages a DAG-based consensus on the main chain to ensure availability and high throughput, enhancing transaction scalability through parallel execution across local chains and a simple scheduling technique.
    \item PyloChain addresses the full member auditing problem by externalizing fine-grained semantics of main chain operations.
    \item We evaluate PyloChain to demonstrate its feasibility.
\end{itemize}

The rest of this paper is organized as follows.
Section \ref{sec:background_relatedwork} provides background and related work.
Section \ref{sec:pylochain_systemoverview} describe the system overview of PyloChain.
Section \ref{sec:protocoldesign} details PyloChain's design.
Section \ref{sec:evaluation} presents evaluation results of PyloChain.
Section \ref{sec:discussionandfuturework} discusses PyloChain with future work, and Section \ref{sec:conclusion} concludes.


\begin{figure}[t]
    \centering
    \includegraphics[scale=0.55]{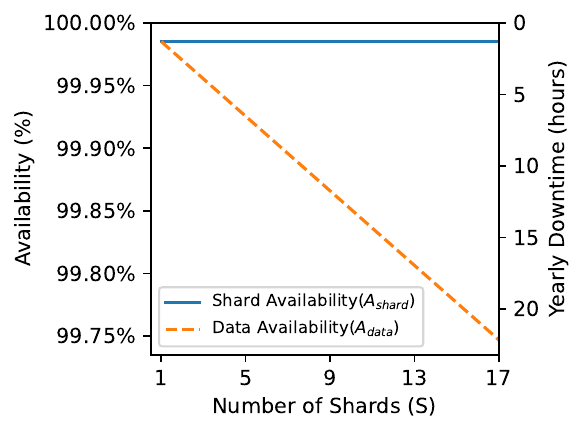}
    \caption{Availability vs. Number of Shards ($N=200$, $p_a=0.995$).}
    \label{pylochain_fig:availability_vs_number_of_shards}
\end{figure}

\section{Background and Related Work}
\label{sec:background_relatedwork}
\subsection{Motivation: Blockchain Sharding and Balanced Approach}
Sharding is regarded as a promising solution to enhance blockchain scalability. For instance, one approach to sharding divides the entire blockchain network into multiple subnetworks, with each subnetwork managing only a portion of the complete blockchain ledger, referred to as a \emph{shard} This division distributes the workload across multiple subnetworks, ultimately reducing consensus, network communication, and storage overhead in each smaller network with fewer members.

However, introducing sharding could reduce the availability of blockchain data, as each sharded data is managed by a smaller number of members.
The failure of a single shard group could lead to the complete unavailability of that shard, causing severe damage to the system. For instance, this could pose a critical issue for applications dealing with business-critical data, such as finance, in consortium blockchains. Consequently, blindly pursuing high scalability through performance sharding may not be a viable option for real-world applications handling highly security-sensitive data (e.g., supply chains on consortium blockchains).

To support our motivation of the necessity of considering availability, we present the following formulation and analyze how sharding affects system availability.
Given $N$ servers, $S$ shards, and annual per-server availability be $p_{a}$, the availability of the entire blockchain dataset be $A_{data}$ is defined as $(A_{shard})^{S}$, where the single shard availability $A_{shard}$ is defined as $A_{\text{shard}} = \sum_{k = 2f+1}^{3f+1}\binom{3f+1}{k}(P_a)^k \, (1-P_a)^{(3f+1)-k}$, which means it follows a binomial distribution, representing the probability that at least $2f_{s}+1$ out of $3f_{s}+1$ servers are available. We set $p_{a}=0.995$ (based on AWS EC2 instances-level's supported monthly availability \cite{AWS_SLA}), $N=200$, and varied $S$ from the non-sharded case ($S=1$) up to $S=17$. The resulting $A_{data}$ and $A_{shard}$ are shown in Fig. \ref{pylochain_fig:availability_vs_number_of_shards}. The results demonstrate that while $A_{shard}$ remains almost constant, $A_{data}$ decreases sharply as the number of shards increases, due to the multiplicative effect of the shard availability. For example, when $S=1$ (no sharding), the expected yearly downtime is approximately 1.3 hours, whereas with $S=17$ shards, it increases to 22.1 hours.


Recent incidents illustrate availability risks even in non-sharded blockchains: Polygon (14 hour outages in 2024), Solana (17-hour halt in 2021, multiple outages in 2022), TON (6-hour disruption in 2024) and Sui (2 hour halt in 2024) experienced failures despite large validator sets \cite{PolygonZkEVMDown, SolanaOutages, SuiOutage, TONOutage}. Sharding amplifies these risks since single shard failures can jeopardize entire dataset access. Considering that these networks collectively secure digital assets worth tens of billions of dollars and operate under constant adversarial conditions, sharded blockchains must treat availability as a first-class security property under perpetual threat.

Therefore, we argue that building a practical sharded blockchain requires balancing both performance and availability. To achieve this, it is necessary to explore trade-offs between traditional performance sharding and availability sharding. As a result, we propose a hierarchical sharded blockchain that divides the network into a two-level structure: the higher level maintains a full copy of all shards, while the lower level holds only a single copy of each shard. This hierarchical design provides excellent opportunities to effectively explore a balanced approach.

\subsection{Sharding Schemes}
We provide existing solutions based on the three sharding schemes to clearly compare our approach.

\subsubsection{Performance Sharding}
Performance sharding is a traditional approach to achieving blockchain scalability due to its maximal parallelism and has therefore been widely adopted in various blockchain systems \cite{SharPer, Saguaro, Ziziphus, AHL, Monoxide, BrokerChain, kokoris2018omniledger, rapidchain, SharDag,Benzene,HieraChain}.
A key challenge in performance sharding lies in the cross-shard protocol, which must ensure atomic commitment of cross-shard transactions across involved shards. For example, traditional coordinator-based two-phase commit (2PC) protocols \cite{kokoris2018omniledger, SharPer, Saguaro, Ziziphus, AHL, BrokerChain,Chainspace}, methods that split transactions into multiple sub-transactions \cite{rapidchain, Monoxide}, techniques that move related states to a single shard \cite{rapidchain,Ziziphus}, or a decentralized protocol where all participants communicate directly without a coordinator \cite{SharPer} have been developed to address this complexity. Those cross-shard protocol is known to be notoriously complex and inefficient.
Interestingly, some methods \cite{Saguaro, HieraChain} introduce a hierarchy within performance sharding to further enhance scalability and simplify cross-shard protocols. In these blockchains, lower-level members store only a single shard of their ledger, while higher-level members either maintain a summarized view of their child ledgers, system-wide metadata for tracking locality, or order cross-shard transactions \cite{Saguaro, HieraChain}.
An approach to enhancing cross-shard scalability in performance sharding utilizes a Trusted Execution Environment (TEE) \cite{Benzene}, where a TEE-generated proof enables lightweight cross-shard block validation, allowing remote shard servers to avoid broadcasting entire blocks.

This performance sharding approach, due to the smaller size of each shard group, involves a significant trade-off in availability, despite achieving higher scalability. In contrast, PyloChain adopts a balanced approach: it uses a hierarchical structure in which lower level members can benefit from performance sharding in terms of reduced storage/networking, and parallel executions, while higher level members maintain a full copy of all shards, ensuring availability in the event of sharding zone failures, and simplifying the cross-shard protocol as well. Also, PyloChain does not rely on TEE.

\subsubsection{Availability Sharding}
Availability sharding allows each member to maintain a full copy of all shards, thereby preventing total data loss in the event of a single shard group failure \cite{Meepo, MeepoJournal, PShard, OHIE}. Moreover, it offers the advantage of leveraging collocated sharding within each member for cross-shard transactions, which helps avoid the complex coordination required for isolated shards as in performance sharding.
In Meepo \cite{Meepo, MeepoJournal}, a sharded consortium blockchain, all members maintain a full copy of the shards, enabling higher cross-shard efficiency by internalizing cross-shard transaction processing within a single member while ensuring shard availability. In PShard \cite{PShard}, every member participates in all shards, which consist of a root chain and multiple child chains, and executes cross-shard transactions using a root chain-driven 2PC-style commit protocol. Similarly, OHIE \cite{OHIE} achieves excellent throughput and availability by allowing each member to maintain up to $j$ parallel chains. In synchronous network settings, each member in OHIE periodically derives a sequence of confirmed blocks (SCB) to establish a global total order across chains for their local view of the parallel chains.

However, availability sharding systems impose a significant burden on each member, as they require every member to hold a full copy of all shards. This becomes especially demanding for less-capable members as the number of members and shards grows.

\subsubsection{Balanced Sharding}
There exists a balanced sharding scheme harmonizing the above two approaches, with some members maintaining a full copy of shards and others a single copy.
Pyramid \cite{Pyramid} uses layered sharding with i-shards and b-shards, which hold a single shard and all shards, respectively. To handle cross-shard transactions, Pyramid employs a b-shard-driven 2PC protocol: the b-shard proposes a cross-shard block to i-shards, which accept or reject based on local transaction conflicts. The b-shard then commits or aborts based on the responses. In contrast, PyloChain streamlines this by executing cross-shard transactions on the main chain based on total order and notifying local chains in a single phase. Unlike Pyramid, which supports cryptocurrency applications in a public blockchain, PyloChain is designed for key-value (KV)-based general workloads \cite{AHL,HyperledgerFabric} in a permissioned blockchain.

DyloChain \cite{dylochain} employs a hierarchical sharded model where higher-level M-nodes aggregate and order local blocks for the main chain, validate transactions, and execute cross-shard transactions under an order-execute-order-validate (O-X-O-V) model. The results are then synchronized back into local chains. However, DyloChain relies on a synchronous network and requires a fixed number of local blocks per synchronous round. This dependency can reduce main chain scalability because slower members may delay main block production, causing substantial latency. In addition, as the main block size grows under high concurrency, local transactions are more likely to be aborted by global transactions, increasing synchronization overhead.
PyloChain extends DyloChain to improve shard scalability under a partially synchronous network. First, PyloChain leverages a DAG-based BFT protocol \cite{bullshark} to efficiently order a large number of concurrent local blocks, which increases main chain throughput and reduces delay even when some higher-level members are slower. Second, to reduce the abort rate of local transactions when main block size grows, PyloChain introduces a simple but effective scheduling technique that improves the cross-shard (or global) transaction processing scalability. Additionally, PyloChain requires a fine-grained auditing mechanism to handle higher-level members in the partially synchronous network.


We highlight the differences between existing approaches and PyloChain in Table \ref{table:comparisons}. Note that in the table, the main chain concept is present only in the balanced sharding scheme. We enhance the overall shard scalability by improving both DyloChain's main chain scalability and cross-shard transaction processing scalability.

\begin{table*}[t]
    \centering
    \caption{Differences from Existing Blockchain Sharding Blockchains.}
    \label{table:comparisons}
    \begin{tabular}{l|c|c|c|c|c|c|c}
        \toprule
                                     & Sharding          & Shard         & Main Chain             & Main Chain    & Cross-Shard TX \\
                                     & Scheme            & Scalability   & Network Model          & Scalability   & Scalability    \\
        \midrule
        Meepo \cite{Meepo}           & Availability      & Low           & -                      & -             & High           \\
        OHIE \cite{OHIE}             & Availability      & Low           & -                      & -             & High           \\
        HieraChain \cite{HieraChain} & Performance       & Very High     & -                      & -             & Low            \\
        Saguaro \cite{Saguaro}       & Performance       & Very High     & -                      & -             & Low            \\
        DyloChain \cite{dylochain}   & Balanced          & Mid.          & Sync.                  & Low           & Mid.           \\
        \textbf{PyloChain}           & \textbf{Balanced} & \textbf{High} & \textbf{Partial Sync.} & \textbf{High} & \textbf{High}  \\
        \bottomrule
    \end{tabular}
\end{table*}
\subsection{Other Scaling Solutions}
Another approach to improving blockchain sharding systems involves resharding techniques \cite{BrokerChain, SharDag, HieraChain, Saguaro, Ziziphus, dylochain}, which aim to reduce the number of costly cross-shard transactions by relocating states across shards. Yet, while resharding techniques can facilitate PyloChain, they are not considered in this paper.
Parallel smart contract execution engines \cite{OptMe, BlockSTM} process transactions using multiple threads within a member after consensus output. In contrast, PyloChain differs by executing transactions first through parallel local chains and performing validation (for local transactions) after consensus.
Stateless blockchain \cite{Porygon} introduces off-chain storage nodes that passively store state to provide data availability, with a separate ordering committee for transaction ordering and execution committees for parallel execution. However, this approach requires multiple rounds of committee-based coordination to support resource-constrained devices, whereas PyloChain employs a hierarchical main chain for immediate finality, enabling efficient synchronization across local chains without such overhead.
Ethereum 2.0's Danksharding \cite{Danksharding} allows layer-2 rollup solutions to post sharded data (called blob) to the layer-1 blockchain in a cost-effective way. This is completely orthogonal to the scalability problem in sharding blockchains that PyloChain addresses.

\section{System Overview}
\label{sec:pylochain_systemoverview}
\subsection{Assumptions and Models}
We assume a permissioned consortium blockchain where the identities of all members are known in advance.
We say members are \emph{Byzantine} if they arbitrarily deviates from the protocol, others we call them \emph{honest}.
A \emph{zone} is a logically partitioned sub-network that independently operates a sharded blockchain.
There exists $f_{L}$ and $f_{F}$ Byzantine failures out of $3f_{L} + 1$ local members within each zone and $3f_{F}+1$ full members across zones, respectively, implying $3f_{L} + 2$ members in a zone.
%
There could be additional machines to be recruited upon detecting the failure of the full member in a zone for recovery.
A zone could become unavailable; 
yet the maximum number of zone failures is limited to $f_{F}$, and the probability of zone failure is mutually independent between zones.
We consider adaptive adversaries capable of corrupting members over time and coordinating collusion attacks. However, the number of corrupted members is bounded by $f_L$ within each zone and $f_F$ across zones for local and full members, respectively.
We assume cross-zone communication is limited to full members; local members do not communicate across zones.
We assume that a PyloChain server participates either as a full member or as a local member within each zone, but not both.

We assume a partially synchronous network where synchronous and asynchronous intervals alternate. In asynchronous intervals, network partitioning may occur. While the duration of such periods is unknown a priori, they eventually terminate within a finite time after the global stabilization time (GST), denoted by $\Delta_{GST}$. Beyond GST, the system remains synchronous for a sufficiently long interval, ensuring that all messages are delivered within $\Delta_{Sync}$.
We assume that cryptographic primitives used in PyloChain are not broken.
We assume that Sybil attacks \cite{SybilAttack} are prevented because all participants are fully authenticated.
Eclipse attacks \cite{EclipseAttack} may occur and can delay messages in an asynchronous way, but under the partially synchronous network assumption, such delays can only last for a finite interval. After GST, message delivery follows a bounded delay, and all messages between participants are eventually delivered.




\subsection{Overview}

\begin{figure}
    \centering
    \includegraphics[scale=.50]{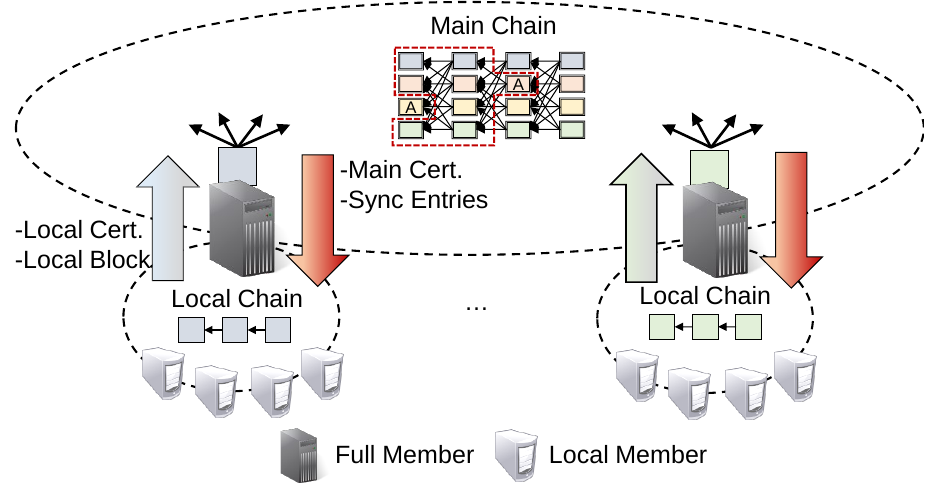}
    \caption{System Overview. PyloChain is composed of multiple lower-level local chains and a single higher-level DAG-based main chain.}
    \label{fig:pylochain_hierarchy_overview}
\end{figure}

PyloChain is a hierarchical blockchain composed of multiple independent sharding zones.
Within each zone, a lower-level sharded local chain is managed by local members, while across zones, a higher-level main chain is managed by full members, as shown in Fig. \ref{fig:pylochain_hierarchy_overview}.
A full member maintains the higher-level DAG-based main chain \cite{narwhaltusk,bullshark} with a full copy of all shards for availability. It is responsible for relaying protocol messages across the hierarchy, including bottom-up messages that relay local blocks with local certificates verifying their correctness to the main chain, and top-down messages that relay main block processing certificates with sync entries containing relevant states (e.g., from aborted local or global transactions) to local members.

A local member maintains the lower-level local chain with a single copy of a shard and, using a PBFT-like protocol \cite{castro1999practical,PeerBFT} with other local members in the same zone, reaches consensus on the local blocks proposed by their respective full member to produce a locally agreed-upon sequence. Local members can verify the state of local transactions but cannot verify global transactions. They also detect potentially faulty full members through a fine-grained auditing mechanism and, upon detecting failure, can replace the faulty full member with a newly recruited honest one.

Clients submit incoming user transactions to a member within the same zone operating a PyloChain server that they trust.
Each user transaction includes read/write sets for a state database, where each entry contains a (key, value, version) tuple, with the version being defined by a block number and transaction offset \cite{HyperledgerFabric}.
Each state is associated with ownership information (i.e., a shard index), which is used to determine the transaction type.

\paragraph*{Local Transaction and Global Transaction}
PyloChain uses the O-X-O-V model \cite{dylochain}, categorizing transaction types into local (single shard) and global (multiple shards). We summarize the overall transaction processing flow as follows.
Local transactions are executed in parallel across multiple local chains, speculatively updating local states, while global transactions are executed and validated only on the main chain.
On the main chain, local transactions are validated based on their read/write sets to ensure consistency with the main chain's state, which reflects the committed states of global transactions. So, the global transactions may abort concurrently executed local transactions if they have dependencies. If local transactions are successfully validated, they are committed to the main chain. Otherwise, they are aborted and rolled back in their respective local chains with the latest main chain state. For detailed validation mechanism across local transactions and global transactions, we refer to interference handling and version map techniques in \cite{dylochain}.

\begin{figure*}[t]
    \centering
    \includegraphics[scale=0.50]{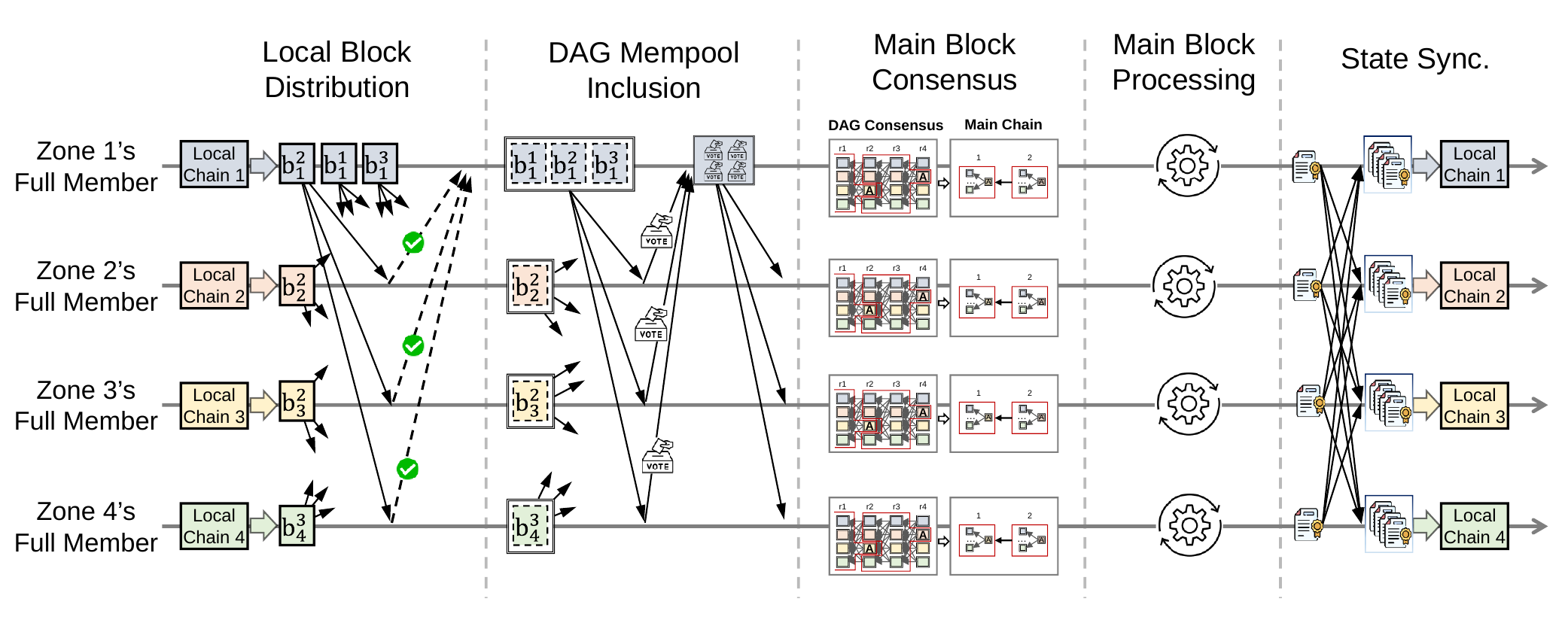}
    \caption{The Main Chain Protocol of PyloChain. Within each sharding zone, full members concurrently distribute their local blocks to ensure availability. They then propose an ordering for these blocks—while preserving their local sequence—for inclusion in the DAG mempool. Subsequently, once a main block is generated by a consensus, each full member processes the main block. Then, the processing results are exchanged among the full members, forming a certificate signed by a quorum of full members before being delivered to the local chains.}
    \label{fig:pylochain_main_chain_protocol} 
\end{figure*}

\section{PyloChain: The Protocol Design}
\label{sec:protocoldesign}
This section provides the protocol design of PyloChain.

\subsection{Local Chain}
Local chains are sharded blockchains managed independently within each zone, where full members and local members reach BFT consensus independently from other zones. The primary responsibility of local chains is the speculative execution of local transactions, significantly reducing processing overhead on the main chain. Local chains may need to revert their states if local transactions are finally aborted on the main chain due to interference or if global transactions require updating local chain states.

We describe operations of local chains as follows.
A full member receives transaction requests from clients within the same zone, checks its validity (e.g., signature and invoking smart contract's arguments), and categorizes each transaction type based on their accessed states (e.g., through simulation or static analysis).
For local transactions, the full member executes the transaction, calculates the read/write set, and updates the local state speculatively.
For global transactions, the full member does not execute them and simply includes them in the next local block.
Subsequently, the full member generates a local block that includes the aforementioned user transactions and broadcasts it to the local members in the same zone. The local members use their local BFT consensus protocol (e.g., \cite{castro1999practical}) to verify the integrity of the full member's local block proposal. They check whether the full member has correctly executed the local transactions and calculated a valid read/write set. If validated, each local member speculatively updates its corresponding local state and confirms the agreed-upon sequence of the proposed local block in their local chain by submitting endorsements, resulting in a local commit certificate for the block.
When a full member becomes behaves maliciously within a zone, PyloChain simply follows a PBFT-like view-change mechanism within the same zone.


\subsection{Main Chain}
\label{pylochain:main_chain_protocol}
Once a local block is committed within a zone, each full member initiates the main chain protocol as illustrated in Fig. \ref{fig:pylochain_main_chain_protocol}, to commit the local block at the main chain. We now describe the detailed phase-by-phase operations of the main chain protocol.

\paragraph{Local Block Distribution}
The goal of \emph{local block distribution} phase is to ensure the availability of local blocks. Specifically, full members collect commit certificates for their local blocks and then reliably broadcast these blocks to other full members, guaranteeing local block availability by receiving $2f_{F}+1$ acknowledgments from other full members. For efficiency, full members effectively leverage a parallel workers architecture \cite{narwhaltusk} to concurrently distribute local blocks.
While such concurrent distribution of local blocks provides performance benefits by maximizing network bandwidth utilization, it is highly likely that the locally agreed sequence of local blocks will not be preserved in the final consensus order.
For example, if there are two local blocks, $b^{1}_{1}$ and $b^{2}_{1}$, from zone $1$, $b^{2}_{1}$ might complete its broadcast before $b^{1}_{1}$ due to network conditions. As a result, $b^{2}_{1}$ can appear before $b^{1}_{1}$ in the final order, which violates the locally agreed sequence. We address this issue in the next DAG mempool inclusion phase.


\paragraph{DAG Mempool Inclusion}
The purpose of \emph{DAG Mempool Inclusion} phase is to include the digest of $2f_{F}+1$ availability-guaranteed local blocks, into a vertex, and incorporate it into the DAG mempool. This phase consists of two main operations:
First, local order-aware vertex creation: the vertex to be proposed is created by each full member, in a way that it maintains the locally agreed sequence of concurrently broadcasted local blocks. For this, each full member proposes the local block by matching its number in one-to-one correspondence with a monotonically increasing round number as defined by the DAG mempool.
This ensures that even if local blocks are completed in different orders during the local block distribution phase, their local sequence is preserved in the DAG mempool seamlessly, with no gaps.
Second, DAG mempool inclusion: The full member proposes the created vertex to other full members and obtains $2f_{F}+1$ votes to finally produce a DAG mempool inclusion certificate. This certificate is then re-broadcast so that other full members incorporate the vertex into their respective local views of the DAG mempool. For details on the process, including round advance rule, voting rule and commit rule, refer to \cite{narwhaltusk, bullshark}.

Note that although proposing only one local block per round is conceptually simple, but it may lead to performance issues, as the local block corresponding to a particular round number might be completed much later, while subsequent local blocks complete much earlier. In such cases, idle time increases, and communication overhead can surge dramatically. To address this, PyloChain can apply a simple optimization allowing multiple monotonically increasing local blocks to be simultaneously proposed in a single vertex.
Since each vertex proposal only includes a list of small-sized block digests, the performance penalty in terms of network bandwidth is negligible.

\paragraph{Main Block Consensus}
For local blocks included in the DAG mempool, full members periodically apply a locally executed deterministic BFT algorithm, reaching consensus on a committed sub-DAG, i.e., main block.
To achieve this, we follow the partially synchronous Bullshark \cite{bullshark} protocol, summarized as follows: In every even-numbered round, each full member selects a predefined anchor vertex (e.g., in a round-robin manner) and includes in the sub-DAG all vertices within the causal history \cite{bullshark} of the anchor, specifically those between the current anchor and the previous anchor vertex.
The vertices included in the sub-DAG are then arranged according to some deterministic topological ordering logic (e.g., depth-first search) to establish a total order across local chains.
For example, in Fig. \ref{fig:pylochain_main_chain_protocol}, the anchor vertices are denoted by "A" in the second and fourth rounds. The vertices between these two anchors form the sub-DAG, indicated by the red dotted line. The main block, which includes all vertices in the sub-DAG, outputs the corresponding local blocks' digests, the block number, its block hash, and the previous hash.
In summary, by performing consensus locally and using a single certificate for the anchor block to commit multiple mempool blocks—i.e., all of its causal histories—the main chain can produce a communication-efficient and high-throughput main block.

Note that DAG consensus maintains an in-memory data structure whose size increases with round numbers, which cannot realistically be contained entirely within bounded memory. Therefore, periodic garbage collection is necessary, but it may introduce the risk of dropping certain vertices (i.e., fairness issues) in an asynchronous network \cite{narwhaltusk}. PyloChain assumes a partially synchronous network model \cite{bullshark}, so all messages arrive within a bounded time after GST and are included in the DAG before being garbage collected.
Alternatively, to relax this assumption for practicality, DAG structures marked for garbage collection can be periodically offloaded to an external storage service, such as a layer-2 storage network.




\paragraph{Main Block Processing}
After the main block consensus outputs a main block, i.e., an ordered list of local blocks from the committed sub-DAG, the full member starts to process all transactions within the main block.
For main block processing, the full member maintains the main chain-level state DB and version map \cite{dylochain}, which enables consistent validation across local chains.
For local transactions, since they have already been executed in the local chain, they contain a read/write set. The full member identifies this read/write set and performs validation based on multi-version concurrency control (MVCC) \cite{HyperledgerFabric}. If validation succeeds, the write set is applied to both the state DB and the version map accordingly.
For global transactions, they are executed in the order determined by the main block, calculating their read/write sets, which are then applied to the state DB as well as the version map.
The processing results of the main block include aborted local transactions and all global transactions with their corresponding main chain states. Each full member appends these results to the sync entries, which are then delivered to the respective local chains.


However, one issue is that speculatively executed local transactions can be aborted by global transactions on the main chain. Such aborts increase the size of sync entries and reduce the number of meaningful committed transactions, negatively impacting overall performance.
To address this, PyloChain applies a simple yet effective scheduling technique: The full member filters global transactions to process all local transactions first, before handling the global transactions during main block processing.
This ensures that global transactions within a main block do not interfere with local transactions in the same main block.
We consider this effect quite significant, as a DAG-based main block with its throughput-oriented and higher concurrency, can include a very large volume of transactions from multiple local chains, thereby avoiding many local transactions being aborted by global transactions.

\begin{figure}[t]
    \centering
    \includegraphics[scale=.9]{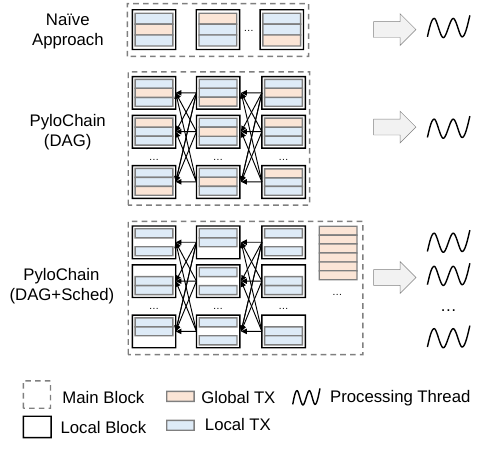}
    \caption{Comparisons of main block processing. A naive approach simply aggregates a fixed number of local blocks from each local chain into a single main block, resulting in a linearly growing chain with a single processing thread. In contrast, PyloChain (DAG) concurrently appends local blocks from each local chain, and PyloChain (DAG+Sched) further optimizes processing with multiple threads and scheduling.} 
    \label{fig:pylochain_executions} 
\end{figure}

Additionally, this scheduling technique offers an opportunity to process mutually independent local blocks from different local chains in parallel, as all global transactions are filtered to the end. This can boost the processing speed of each main block, especially as the number of zones increases.
For comparison, we illustrate our processing methods in Fig. \ref{fig:pylochain_executions}, showing the DAG-only PyloChain and the scheduling-enhanced version of PyloChain, denoted as PyloChain (DAG) and PyloChain (DAG+Sched), respectively, alongside a naive baseline approach that simply collects a fixed number of local blocks from each local chain (e.g., \cite{dylochain}).

We briefly discuss scheduling considerations. Reordering transactions within a main block does not affect fairness in user-perceived commit latency, because commit events are emitted only after the entire main block is processed. In terms of abortability, global scheduling prevents interference among local transactions within a block, and any interference across blocks is independent of scheduling, so fairness is preserved. The scheduling overhead is low, as it simply filters global transactions by iterating over blocks; this step can also be parallelized to reduce latency. A potential fairness concern may arise only outside the protocol, if a PyloChain server operator monitors execution logs in real time and externally discloses results before main-block completion. Finally, scheduling does not affect state consistency. Although global transactions appear in local blocks, they are neither executed nor applied to local states; their states are finalized only during main-block processing. Thus, reordering them within a block does not impact local-chain states.

\paragraph{State Synchronization}
The purpose of the final phase of the protocol is that the results of main block processing are reliably propagated to the respective local chains.
After main block processing, each full member generates a main block processing proof for the main block, summarizing the results of main block processing, and broadcasts it among themselves to collect $f_{F}+1$ proofs. The proof includes the main block number, main block hash, previous main block hash, the sequence of local block headers within the main block, the identity of the creator, and the digest values of the sync entries that need to be committed to each local chain. Each sync entry consists of a key, value, and transaction ID. Once $f_{F}+1$ proofs are gathered, each full member creates a main block processing certificate and propagates it along with the payloads of the sync entries, to their respective local chain.
Then, the local members verify the main block processing certificate and sync entries, allowing them to reliably confirm that the local blocks from their local chain have been committed to the main chain. Lastly, they safely synchronize the states of aborted local transactions and committed global transactions to the local chain using the payloads contained in the sync entries.

\DontPrintSemicolon
\SetAlgoLined
\SetNoFillComment

\SetKwProg{creation}{on creating}{:}{}
\SetKwProg{event}{on receiving}{:}{}
\SetKwProg{Procedure}{Procedure}{:}{}

\SetKw{Continue}{continue}
\SetKwData{Candidates}{$candidates$}
\SetKwData{Next}{$next$}
\SetKwData{Vertex}{$v$}
\SetKwData{Round}{$r$}
\SetKwData{OrderedDigests}{$\tau$}
\SetKwData{Mempool}{\mathcal{M}}
\SetKwData{MempoolBlocks}{$\mathcal{B}$}

\SetKwData{SyncEntries}{$ents$}

\SetKwProg{Fn}{Function}{:}{}

\SetKwFunction{CreateVertex}{createVertex}
\SetKwFunction{CreateMainBlock}{\textsf{CreateMainBlock}}
\SetKwFunction{ProcessMainBlock}{\textsf{ProcessMainBlock}}
\SetKwFunction{AppendEntry}{\textsf{AppendEntry}}
\SetKwFunction{Validate}{\textsf{Validate}}
\SetKwFunction{Update}{\textsf{Update}}
\SetKwFunction{Execute}{\textsf{Execute}}
\SetKwFunction{GetStates}{\textsf{GetState}}
\SetKwFunction{Ownership}{\textsf{Ownership}}

\SetKwData{Buffer}{buffer}
\SetKwData{BufferPopNext}{buffer.pop(next)}
\newcommand{\mycommentstyle}[1]{\textcolor{blue}{\textit{#1}}}
\SetCommentSty{mycommentstyle}

\begin{algorithm}[t]

    \caption{The Main Chain Protocol}
    \label{alg_mainchain_protocol}
    \small

    \tcp{For Full Members}
    $\mathcal{M}$ $\leftarrow$ $\emptyset$; \label{alg:init} \tcp{Initialize DAG mempool}
    \tcp{Phase 1: Local Block Distribution}
    \event{\label{alg:BLK_start}\emph{a local block certificate $cert_{local}$ for $b_{i}^{z}$ at $p \in \Pi_{F}^{z}$}}{
        $\mathcal{M}_{z}$.\textsf{Add}($b_{i}^{z}$); broadcast $\langle$REPL, $z$, $i$, $b_{i}^{z}$$\rangle$$_{\sigma_{p}}$ to $\Pi_{F}$\;
        \label{alg:BLK_end}
    }

    \event{\label{alg:REPL_start}\emph{$\langle$REPL, $z$, $i$, $b_{i}^{z}$$\rangle$$_{\sigma_{p}}$ at $q \in \Pi_{F}$}}{
        $\mathcal{M}_{z}$.\textsf{Add}($b_{i}^{z}$);
        reply $\langle$ACK, $z$, $i$, $D(b_{i}^{z})$$\rangle$$_{\sigma_{q}}$ to $p$
        \label{alg:REPL_end}
    }
    \tcp{Phase 2: DAG Mempool Inclusion}
    \event{\label{alg:AVAIL_start} \emph{$2f_{F}+1$ $\langle$ACK, $z$, $i$, $D(b_{i}^{z})\rangle$$_{\sigma_{p}}$ at $p \in \Pi_{F}^{z}$}}{

        \If{$\mathcal{M}_{z}$.\textsf{\emph{Next()}} = $i$}{
            \Candidates $\leftarrow \emptyset$ \;
            \While{$b \leftarrow$ $\mathcal{M}_{z}$.\emph{\textsf{Pop()}} \textbf{is not} nil}{
                \Candidates $\leftarrow$ $\Candidates$ $\cup$ $\{D(b)\}$\;
            }
        }\label{alg:monotonic_ordering_end}

        $cert_{avail}$ $\leftarrow$ a certificate from $2f_{F}+1$ ACK msgs. \;
        $DAG$.\textsf{Include}($candidates$, $cert_{avail}$) \;
        broadcast $\langle$AVAIL, $cert_{avail}$, $z$, $i$$\rangle$$_{\sigma_{p}}$ to $\Pi_{L}^{z}$\;
        \label{alg:AVAIL_end}
    }
    \tcp{Phase 3: Main Block Consensus}
    \event{\label{alg:main_block_processing_start} $\langle$\emph{CON}, \OrderedDigests, $j$$\rangle$ from $DAG$.\textsf{\emph{Consensus()}} at $p \in \Pi_{F}^{z}$} {
        $b_{m}^{j}$ $\leftarrow$ $\CreateMainBlock(\mathcal{M}, \OrderedDigests, j)$ \;
        \tcp{Phase 4: Main Block Processing}
        \SyncEntries $\leftarrow$ $\ProcessMainBlock(b_{m}^{j})$ \;
        \tcp{Phase 5: State Synchronization}
        $hdr$ $\leftarrow$ \textsf{BlockHeader}($b_{m}^{j}$) \;
        $\pi_{j}$ $\leftarrow$ \textsf{MainBlockProof}(\SyncEntries, $hdr$, $j$) \;
        $cert_{proc}$ $\leftarrow$ exchange $\pi_{j}$ among $\Pi_{M}$ \;
        broadcast $\langle$PROC, $cert_{proc}$, $j$, $\SyncEntries_{z}$$\rangle$$_{\sigma_{p}}$ to $\Pi_{L}^{z}$\;
        \label{alg:main_block_processing_end}
    }

    \tcp{For Local Members}
    \event{\label{alg:local_BLK_start}\emph{a local block certificate} $cert_{local}$ for \emph{$b_{i}^{z}$ at $p \in \Pi_{L}^{z}$}}{
        verify $cert_{local}$ and start a timer $\Delta_{avail}$
        \label{alg:local_BLK_end}
    }

    \event{\label{alg:avail_processing_start} $\langle$\emph{AVAIL}, $cert_{avail}$, $z$, $i$$\rangle$$_{\sigma_{p}}$ at $p \in \Pi_{L}^{z}$} {
        verify $cert_{avail}$, exit the timer $\Delta_{avail}$, and start a timer $\Delta_{proc}$
        \label{alg:avail_processing_end}
    }
    \event{\label{alg:sync_processing_start} $\langle$\emph{PROC}, $cert_{proc}$, $j$, $\SyncEntries_{z}$$\rangle$$_{\sigma_{p}}$ at $p \in \Pi_{L}^{z}$} {
        verify $cert_{proc}$ and exit a timer $\Delta_{proc}$\;
        update sync entries in $ent_{z}$, according to $j$, to local state DB
        \label{alg:sync_processing_end}
    }



\end{algorithm}

\begin{algorithm}[t]

    \caption{Main Block Processing}
    \small


    \label{alg_mainblockprocessing}
    \Fn{\ProcessMainBlock{$b_{m}^{j}$}} {
        \tcp{Schedule global TXs at the end}
        $localTXs$ $\leftarrow$ extract local TXs from $b_{m}^{j}$ \label{alg:mbp_extract_local_txs} \;
        $globalTXs$ $\leftarrow$ extract global TXs from $b_{m}^{j}$ \label{alg:mbp_extract_global_txs} \;
        \SyncEntries $\leftarrow$ \{\} \label{alg:mbp_init_syncent}\;
        \tcp{Process local TXs}
        \ForEach{$tx \in localTXs$}{ \label{alg:mbp_process_localtx_start}
            $res$ $\leftarrow$ $\Validate(tx)$\;
            \lIf{$res$ = $success$}{
                $\Update(tx)$; \Continue
            }
            \AppendEntry{$\SyncEntries$, $tx.Id$, $tx.WSet$} \;

            \label{alg:mbp_process_localtx_end}
        }

        \tcp{Process global TXs}
        \ForEach{$tx \in globalTXs$}{\label{alg:mbp_process_globaltx_start}
            $rwset$ $\leftarrow$ $\Execute(tx)$ \;
            \AppendEntry{$\SyncEntries$, $tx.Id$, $rwset.WSet$} \;
            $\Update(tx)$ \;
            \label{alg:mbp_process_globaltx_end}
        }
        \KwRet \SyncEntries \label{alg:mbp_process_return}\;
    }
    \Fn{\AppendEntry{$ents$, $txid$, $wset$}} {\label{alg:mbp_append_start}
        \ForEach{$w \in wset$}{
            $z$ $\leftarrow$ $\Ownership(w.key)$ \;
            $v_{lat}$ $\leftarrow$  $\GetStates(z, w.key)$\;
            \textsf{Add}$(\SyncEntries_{z}, \{txid, w.key, v_{lat}\})$
        }\label{alg:mbp_append_end}
    }

\end{algorithm}

\subsection{Algorithms}
We provide the pseudocode of the main chain protocol of PyloChain in Algorithm \ref{alg_mainchain_protocol} and Algorithm \ref{alg_mainblockprocessing} to clearly illustrate its operations.
We introduce additional notations as follows.
A mempool is denoted by $\mathcal{M}$, and the mempool for zone $z$ by $\mathcal{M}_{z}$.
$\mathcal{M}_{z}$ adds zone $z$'s local blocks with the method \textsf{Add()}, which stores the block and internally increments the latest monotonically increasing local block number. It supports \textsf{Pop()}, which returns and deletes the lowest value among the stored monotonically increasing sequences, and \textsf{Next()}, which provides the next value following the most recently returned value.
Full and local members from zone $z$ are denoted by $\Pi_{F}^{z}$ and $\Pi_{L}^{z}$, respectively.
$\Pi_{F}$ denotes all full members across zones.
A signed message by participant $p$ is denoted as $\langle\cdot\rangle_{\sigma_{p}}$.
We define $DAG$ as a black-box module that performs DAG mempool inclusion and main block consensus, providing two functions: \textsf{Include} and \textsf{Consensus}. The \textsf{Include} function adds a newly created vertex to the DAG, while the \textsf{Consensus} function interprets the DAG and notifies the topologically ordered committed sub-DAG with a sequence number for the main chain. We first describe the protocol for full members and then explain the protocol for local members.

Initially, each full member maintains an empty mempool, $\mathcal{M}$ (L\ref{alg:init}). Upon receiving a local block certificate for a local block $b_{i}^{z}$ from $\Pi_{L}^{z}$, a full member $p \in \Pi_{F}^{z}$ adds the block to its mempool $\mathcal{M}$ and broadcasts $\langle$REPL, $z$, $i$, $b_{i}^{z}$$\rangle$$_{\sigma_{p}}$ to other full members for block availability (L\ref{alg:BLK_start}-L\ref{alg:BLK_end}). Upon receiving the REPL message, a full member stores the local block from zone $z$ into its $z$-indexed mempool $\mathcal{M}_{z}$ and replies with $\langle$ACK, $z$, $i$, $D(b_{i}^{z})$$\rangle$$_{\sigma_{p}}$ (L\ref{alg:REPL_start}-L\ref{alg:REPL_end}). When the proposing member $p$ has collected $2f_{F}+1$ ACK messages including its own (L\ref{alg:AVAIL_start}), it prepares to propose a vertex containing the set of block digests. The \textsf{Next()} method of $\mathcal{M}_{z}$ returns the next block number to be proposed from zone $z$. If this matches the block number $i$ from the ACK messages, member $p$ consecutively executes $\textsf{Pop()}$ from $\mathcal{M}_{z}$, adding digests $D(b)$ to $candidates$ from block number $i$ up to the latest available block number. The member $p$ then generates a block availability certificate $cert_{avail}$ from the $2f_{F}+1$ ACK messages, asynchronously executes $DAG$.\textsf{Include} with these digests, and broadcasts $\langle$AVAIL, $cert_{avail}$, $z$, $i$$\rangle$$_{\sigma_{p}}$ to local members in zone $z$ (L\ref{alg:AVAIL_start}-L\ref{alg:AVAIL_end}).

        The asynchronously executed result from $DAG$.\textsf{Consensus()} is returned within a CON message as event notifications with ordered digests $\OrderedDigests$ and their sequence number $j$. Using these, the $j$-th main block $b_{m}^{j}$ is generated by \textsf{CreateMainBlock}, processed by \textsf{ProcessMainBlock} (which will be explained in \ref{alg_mainblockprocessing}), returning the updated states as $ents$. A processing proof, which is a signature over inputs including entries $ents$, local block headers $b_{m}^{j}.Hdrs$, and the main block number $j$, is generated by \textsf{MainBlockProof} and then exchanged among full members to create a processing certificate. Note that $ents$ contain both zone-level hashes and a global hash, and the zone-level hashes can be indexed.
        Finally, each full member broadcasts the main chain processing results and zone-specific entries $ent_{z}$ within the message $\langle$PROC, $cert_{proc}$, $j$, $ents_{z}$$\rangle$$_{\sigma_{p}}$ to all local members in zone $z$, with invoking local chain consnesus (L\ref{alg:main_block_processing_start}-L\ref{alg:main_block_processing_end}).


        The main block processing is presented in Algorithm \ref{alg_mainblockprocessing}.
        The algorithm takes the main block $B$, extracts global transactions through filtering (L\ref{alg:mbp_extract_global_txs}), and initializes sync entries (L\ref{alg:mbp_init_syncent}).
        Then, it performs MVCC-based validation \cite{dylochain} using \textsf{Validate} on the local transactions from the filtered main block. If validation succeeds, the results are updated in the main-chain state DB (\textsf{Update}), and the protocol proceeds to the next iteration. If not, i.e., when speculative local updates are aborted, the latest main-chain states of the involved write keys must be synchronized to the local state, for which we use \textsf{AppendEntry} (L\ref{alg:mbp_process_localtx_start}–\ref{alg:mbp_process_localtx_end}).
        The \textsf{AppendEntry} operation adds a synchronization entry to the given $ents$. Specifically, for each write $w$ in the write set $wset$, it identifies the ownership $z$ (i.e., shard index) of its key $w.key$, retrieves the latest state $v_{lat}$ from the state DB, and appends it to $ents$ (L\ref{alg:mbp_append_start}–L\ref{alg:mbp_append_end}).
        Subsequently, global transactions are executed sequentially to compute their $rwset$, update the results to the state DB, and append them to $ents$ (L\ref{alg:mbp_process_globaltx_start}–\ref{alg:mbp_process_globaltx_end}).
        After processing global transactions, the algorithm finally returns the sync entries $ents$ (L\ref{alg:mbp_process_return}).

        We describe the local member protocol as follows: After a local member $p$ in a zone $z$ identifies the local block certificate $cert_{local}$ for $b_{i}^{z}$, the local member verifies its certificate and starts a timer $\Delta_{avail}$ (L\ref{alg:local_BLK_start}-\ref{alg:local_BLK_end}). When an AVAIL message is delivered by the full member in $z$, each local member verifies the $cert_{avail}$, exits the timer $\Delta_{avail}$, and starts another timer $\Delta_{proc}$ to measure the delay until $b_{i}^{z}$ is processed on the main chain (L\ref{alg:avail_processing_start}-\ref{alg:avail_processing_end}). Finally, upon receiving the corresponding PROC message, each local member verifies the $cert_{proc}$, exits the timer $\Delta_{proc}$, and updates the entries in $ent_{z}$ to the local state database according to $j$ (L\ref{alg:sync_processing_start}-\ref{alg:sync_processing_end}). Note that we intentionally omit the local sequence number for simplicity. Local members within zone $z$ consistently update the entries through the local chain consensus.
        We discuss timer configurations under a partially synchronous network, along with failure handling scenarios, in Section \ref{pylochain:auditing_full_member_trustworthiness} and Section \ref{pylochain:fault_handling}.

        \subsection{Auditing Full Member's Behavior}
        \label{pylochain:auditing_full_member_trustworthiness}
        We now describe the mechanism for auditing full members' trustworthiness for their behaviors.
        Full members, who relay protocol messages between local chains and a main chain, introduce significant security challenges for PyloChain. Malicious relay attacks are classified into two types:
        (1) Bottom-up attacks occur when a full member withholds local blocks (compromising availability) or broadcasts malicious blocks (compromising integrity).
        (2) Top-down attacks occur when a full member fails to deliver main block processing certificates to the local chain (compromising liveness) or reports fake certificates (compromising integrity).



        Among these, data integrity is straightforward to guarantee, as local blocks and main block processing certificates contain signatures from $2f_{L}+1$ local members and $2f_{F}+1$ full members, respectively.
        In other words, even if a malicious full member attempts to inject fake local blocks into the DAG mempool, these blocks will not be included in the honest full members' DAG mempool without valid local block certificates. Similarly, fake entries generated by malicious full members through incorrect main block processing will not be applied to the local chain without valid full members' certificates.
        While a full member might reorder local blocks in the DAG mempool against the locally agreed sequence, local members can detect this by verifying the order of local block headers in the subsequent main block processing certificate.


        But timing-related faults, such as delaying or dropping relayed messages, require careful handling. Local members should be able to confirm that their local blocks are included in the main chain in a timely manner via the main block processing certificate from their full member. However, because the main chain operates in a partially synchronous network and is managed solely by full members, local members lack direct visibility into its status. Thus, the full member must provide proof of correct behavior to the local members.

        PyloChain addresses this by making each full member responsible for externalizing fine-grained main chain operations, along with their corresponding certificates, to its local members, including local block availability, DAG mempool inclusion, and main block consensus and processing.
        This allows local members to use a timer to check whether each main chain operation is delivered within the expected latency once a local block is agreed upon in their local chain. We first define the expected maximum latency for these operations during the normal synchronous period.

        For the local block availability certificate, it takes $3\Delta_{Sync}$ (i.e., $\Delta_{avail}$): $1\Delta_{Sync}$ for broadcasting the local block to the full members, $1\Delta_{Sync}$ for collecting $2f_F+1$ acknowledgments from other full members, and $1\Delta_{Sync}$ for delivering these acknowledgments to the local members.
        For the DAG mempool inclusion certificate, it takes $3\Delta_{Sync}$: $1\Delta_{Sync}$ for the full member's vertex proposal, $1\Delta_{Sync}$ for collecting $2f_F+1$ votes from other full members, and $1\Delta_{Sync}$ for delivering these votes to the local members.
        For the main block consensus certificate, it takes $4\Delta_{Sync}$ because consensus is reached every two rounds using the partially synchronous Bullshark algorithm \cite{bullshark}, and the next round's vertex must follow the previous two DAG mempool steps.
        For the main block processing certificate (i.e., $\Delta_{proc}$), it takes $2\Delta_{Sync}$: $1\Delta_{Sync}$ for full members to broadcast their main block processing certificates and gather $f_{F}+1$ certificates, and $1\Delta_{Sync}$ for delivering them to the local members.
        Since the main block consensus is performed locally, it can be combined with the main block processing certificates and delivered in a single batch, with the combined operation expected to take a total of $6\Delta_{Sync}$.
        Note that we omitted externalizing the DAG mempool inclusion certificate in Algorithm \ref{alg_mainchain_protocol} for simplicity, as this phase's certificate can be incorporated into the main block processing certificate.

        By using timers based on the expected confirmation latency of externalized operations, local members gain visibility into main chain operations conducted by full members, allowing them to audit their trustworthiness. However, in a partially synchronous network, unexpected latency periods can occur, during which necessary certificates might not be delivered within the expected confirmation latency.
        In such cases, local members may ask, \emph{``Is this due to being in the GST period, or is it our full member's fault?''} For this, they can either conservatively wait up to $\Delta_{GST}$ for the message to arrive, concluding a fault by the full member if it still doesn't arrive, or proactively determine a fault as soon as the synchronous timer expires.
        Note that the $\Delta_{GST}$ is a theoretical value. In practice, it can be implemented using adaptively increasing timers, e.g., $2\Delta_{Sync}$, $4\Delta_{Sync}$, $8\Delta_{Sync}$, and so on, which reflects the partially synchronous network assumption in which $\Delta_{GST}$ eventually converges to some finite upper bound.
        In PyloChain, we adopt the conservative approach, based on the assumption that Byzantine attacks by full members are rare and are handled by a reputable service provider.

        We describe PyloChain's behavior under asynchronous network conditions during the GST period. In the first case, where messages are eventually delivered after the GST period, local members record uncommitted local blocks in a pending block list during the GST period. Once the GST period ends and the synchronous period begins, missing certificates for the pending local blocks are delivered by the full members, and the corresponding local blocks are removed from the pending list, after which normal operations resume. In the second case, where messages are not received even after the GST period ends, local members also record uncommitted local blocks in the pending block list during the GST period. However, if the expected messages are still not received after the GST period ends and the synchronous period begins, local members conclude that their full member is malicious.


        \begin{algorithm}[t]
            \caption{Full Member Recovery on Main Chain}
            \label{alg:recovery}
            \small

            Assume a new full member $q' \in \Pi^{z}_{F}$ is elected in zone $z$ and joins the main chain network \label{alg:recovery:assume} \;

            $q'$ downloads main chain state from $f_F+1$ full members in $\Pi_F$ \label{alg:recovery_state_catchup} \;
            $q'$ identifies the highest local block number $h$ from $\mathcal{M}_{z}$ \label{alg:recovery:highest_local_block} \;
            $q'$ proposes pending local blocks $\{b^z_i : i > h\}$ to the main chain protocol by resuming normal operations in the local chain \label{alg:recovery:propose_pending_blocks} \;

        \end{algorithm}

        \subsection{Faulty Full Member Handling}
        \label{pylochain:fault_handling}
        Since the main chain of PyloChain leverages a DAG mempool, $2f_{F}+1$ local block availability is supported as long as the local blocks are included in the DAG mempool.
        However, full members play a critical role, as they participate in both local chain and main chain consensus. Therefore, if a full member is malicious, it can impact the system's performance and security.

        We first explain the impact of a faulty full member on the main chain. DAG-based main chain consensus inherently does not require a traditional view change mechanism due to the nature of DAG construction \cite{PartialSyncBullshark,bullshark}. This is because, in DAG mempool construction, no entity is assigned a special role. Even in the presence of a faulty full member, honest full members can proceed completely asynchronously at network speed. Specifically, for local block distribution that ensures data availability, only acknowledgments from $2f_{F}+1$ members are required to proceed.
        On the other hand, the Bullshark consensus proceeds by interpreting DAG construction on a round-by-round basis. In this scheme, every even-numbered round relies on a predetermined anchor vertex (known to all via a deterministic mapping function, e.g., round-robin) to maintain consistency in DAG ordering logic. In scenarios where the anchor vertex belongs to a faulty full member, this member can either: 1) fail to propose the vertex entirely or 2) propose it selectively to only a subset of members. Nevertheless, Bullshark safely skips these faulty vertices by leveraging the $f_{F}+1$-commit rule. By committing the next honest anchor vertex in a subsequent round, Bullshark can retroactively and amortizedly commit the path from that honest anchor vertex back to the previous committed anchor vertex. Consequently, although a faulty full member can slightly delay the consensus speed, it does not negatively impact throughput.This characteristic is a primary advantage of introducing DAG-based consensus into the upper layer of PyloChain, significantly minimizing overhead from malicious actions by full members.

        We now explain how PyloChain recovers a faulty full member.
        Within a zone, a traditional PBFT-like view change mechanism \cite{castro1999practical} is leveraged to replace the faulty full member with a new full member. This mechanism enables the correct migration of the local chain from the previous view to the new view within the zone.
        However, in a hierarchical blockchain, such a traditional view change alone is insufficient, as the main chain must also be considered. We describe the recovery from the main chain perspective as in the Algorithm \ref{alg:recovery}:
        First, the newly elected full member $q'$ within zone $z$ joins the main chain network (L\ref{alg:recovery:assume}). For this, the new full member must obtain participation permission, which can be managed via a consensus-based membership change protocol \cite{dyno}.
        Next, the new full member catches up with the main chain by contacting honest members and downloading DAG mempool and sub-DAGs (L\ref{alg:recovery_state_catchup}). To avoid costly re-execution, the main states can be downloaded as well. For fault tolerance, a new full member first verifies state integrity by querying digests from $f_{F}+1$ members, and then downloads the actual large-volume data from only one member to verify it.
        Then, the new full member identifies the highest local block number whose availability has been ensured on the main chain (L\ref{alg:recovery:highest_local_block}).
        Next, the new full member identifies any pending local blocks that have been locally committed but not yet included in the DAG mempool and  resumes the main chain protocol by starting to submit the pending local blocks to the DAG mempool. Simultaneously, the new full member resumes normal local chain operations by processing incoming client transactions (L\ref{alg:recovery:propose_pending_blocks}).

        \subsection{An optimization}
        PyloChain's main chain requires each full member to broadcast the local blocks of its respective zone to all other full members across zones to support data availability. However, this can rapidly consume network resources as the number of zones, the size of local blocks, and their generation speed increase. To address this, erasure coding techniques (e.g., Reed-Solomon ($3f_{F}+1$, $2f_{F}+1$)) can be utilized for full members \cite{CodedBlockchain24} when they propagate local blocks to other full members. By adopting this coding technique, only a fraction of each encoded local block—rather than the entire block—needs to be propagated, significantly reducing network communication and storage overhead. This would be particularly effective when a high degree of the workload exhibits locality \cite{dylochain}. However, one challenge remains: erasure coding requires costly reconstruction operations to access block contents. Nevertheless, PyloChain is expected to mitigate this cost by performing selective reconstruction only when processing global transactions. We leave the implementation of supporting global transaction processing via such selective reconstruction for future work. Yet, in our evaluation, we demonstrate the potential of this technique under a local-only transaction workload.

        \subsection{Analysis}
        \label{pylochain_properties}
        We analyze the correctness of PyloChain in terms of safety and liveness. We also provide analysis of performance, storage cost, and auditing overhead of PyloChain.
        \subsubsection{Preliminaries}
        \begin{definition}[System Model]
            PyloChain is modeled as the following tuples: $<\Pi, Z, B, S>$ where
            \begin{itemize}
                \item $\Pi$: Set of all members, i.e.,  $\Pi = \Pi_{F} \cup \Pi_{L}$
                      \begin{itemize}
                          \item $\Pi_{F}$: the set of all full members.
                          \item $\Pi_{F}^{z}$: the set of full members within a zone $z$.
                          \item  $\Pi_{L}$: the set of all local members.
                          \item  $\Pi_{L}^{z}$: the set of local members within a zone $z$.
                          \item $\Pi^{z}$: the set of all members within a zone $z$, i.e., $\Pi^{z} = \Pi_{F}^{z} \cup \Pi_{L}^{z}$.
                      \end{itemize}

                \item $Z$: Number of zones.
                \item $B$: Set of blocks in the system, i.e., $B=B_{L} \cup B_{M}$
                      \begin{itemize}
                          \item $b_{z}^{i} \in B_{L}$: the $i$-th local block in zone $z$.

                          \item $b_{m}^{j} \in B_{M}$: the $j$-th main block in the main chain.
                      \end{itemize}
                \item $S$: Blockchain state.
                      \begin{itemize}
                          \item $s_{z}^{i} \subset S$: states to be committed in $b_{z}^{i}$.
                          \item $s_{m}^{j} \subset S$: states to be committed in $b_{m}^{j}$.
                      \end{itemize}
            \end{itemize}
        \end{definition}

        \begin{definition}[Predicates]
            $\quad$
            \begin{itemize}

                \item $\mathsf{commit}_{L}(p, b_{z}^{i})$: a predicate indicating that member $p \in \Pi_{L}^{z}$ has committed the local block $b_{z}^{i}$ to its local chain.

                \item $\mathsf{commit}_{M}(p, b^{j}_{m})$: a predicate indicating that member $p \in \Pi_{F}$ has committed the main block $b^{j}_{m}$ to its main chain.

                \item $\mathsf{contains}(b^{j}_{m}, b_{z}^{i})$: a predicate indicating that the main block $b_{m}^{j}$ contains the $b_{z}^{i}$.

                \item $\mathsf{sync}(p, s^{j}_{m},  s_{z}^{i})$: a predicate indicating that member $p \in \Pi_{L}^{z}$ has synched the resulting states $s_{z}^{i} \in s^{j}_{m}$ from $b^{j}_{m}$ where  $\mathsf{contains}(b^{j}_{m}, b_{z}^{i})$ is true.

            \end{itemize}
        \end{definition}

        \begin{assumption}[Partially Synchronous Network]
            \label{assumption:partial_synchrony}
            The network alternates between synchronous periods (messages delivered within $\Delta_{sync}$) and asynchronous periods (finite but unknown duration $\Delta_{GST}$). After $\Delta_{GST}$, the network remains synchronous for a sufficiently long time.

        \end{assumption}

        \begin{assumption}[Byzantine Resilience]
            \label{assumption:byzantine_resilience}
            The system can tolerate up to $f_{F}$  byzantine faulty full members across zones, and $f_{L}$ byzantine local members within each zone, i.e., $|\Pi_{F}| = 3f_{F}+1, \quad |\Pi_{L}^{z}| = 3f_{L}+1$ for each zone $z$. Accordingly, zonal members for zone $z$ consists of $\Pi^{z} = \Pi_{F}^{z} \cup \Pi_{L}^{z}$, where $|\Pi_{F}^{z}| = 1, \quad |\Pi_{L}^{z}| = 3f_{L}+1$.
        \end{assumption}

        \begin{assumption}[Correctness of Underlying BFT Consensus]
            \label{assumption:correctness_of_underlying_bft}
            The BFT consensus protocols of both the local chain and the main chain guarantee safety and liveness under their respective assumptions.
        \end{assumption}

        \subsubsection{Liveness}
        \label{liveness}

        \begin{lemma}[Eventual Recovery]
            \label{lemma:eventual_recovery}

            If a full member $q \in \Pi_F^z$ fails to timely relay inter-layer messages to $p \in \Pi_L^z$, such misbehavior is eventually detected and recovered by replacing $q$ with a new full member.

        \end{lemma}

        \begin{proof}[Proof]

            Let $p \in \Pi_L^z$ be an honest member holding $\mathsf{commit}_L(p,b_z^i)$.
            After this, $p$ starts timers to monitor main-chain phases (Section \ref{pylochain:auditing_full_member_trustworthiness}).
            If a timer expires before receiving the expected message, the cause may be a faulty $q$ or an asynchronous network period. Local members then initiate a PBFT-like view change to replace $q$ with another member of zone $z$, using the recovery mechanism of Section \ref{pylochain:fault_handling}.
            Liveness of this recovery follows from two assumptions.
            (1) By Assumption \ref{assumption:byzantine_resilience}, at most $f_L$ full members can be faulty among $3f_L+1$, so repeated replacements eventually select an honest member.
            (2) By Assumption \ref{assumption:partial_synchrony}, any asynchronous period is finite, and the network eventually becomes synchronous.
            Thus, an honest full member will eventually be elected during a synchronous period, ensuring the pending messages are eventually relayed.

        \end{proof}


        \begin{lemma}[Eventual Commitment]
            \label{lemma:eventual_main_commit}

            If $\mathsf{commit}_{L}(p,b_z^i)$ holds for $p \in \Pi_L^z$, then a main block $b_m^j$ eventually exists such that $\mathsf{contains}(b_m^j,b_z^i)$ and $\mathsf{commit}_M(q,b_m^j)$ for $q \in \Pi_{F}$ holds.

        \end{lemma}

        \begin{proof}

            Assume $\mathsf{commit}_L(p,b_z^i)$ holds for $p \in \Pi_L^z$.  We consider two cases. \textbf{Case 1:} $q$ is honest and the network is synchronous.  Then $q$ follows Algorithm \ref{alg_mainchain_protocol}.  Local block distribution and DAG inclusion complete within $3\Delta_{sync}$, and the underlying DAG-based BFT consensus for the main chain guarantees liveness by Assumption \ref{assumption:correctness_of_underlying_bft}, which eventually outputs a main block $b_{m}^{j}$ containing $b_z^i$ within $4\Delta_{sync}$. Thus, $\mathsf{commit}_M(q,b_m^j)$ follows. \textbf{Case 2:} $q$ is faulty or the network is asynchronous. Local members may not receive certificates, triggering a timeout.  By Lemma \ref{lemma:eventual_recovery}, recovery process eventually installs an honest full member since faulty members are bounded and asynchrony is finite. Therefore, after a finite time, the PyloChain will reach a state equivalent to Case 1. From this state, $b_z^i$ is eventually included and committed in some main block $b_{m}^{j}$.

        \end{proof}

        \begin{lemma}[Eventual Synchronization]
            \label{lemma:eventual_synchronization}

            If $\mathsf{commit}_M(q,b_m^j)$ holds for $q \in \Pi_{F}^{z}$, then for $b_z^i$ with $\mathsf{contains}(b_m^j,b_z^i)$, $\mathsf{sync}(p,s_m^j,s_z^i)$ eventually holds for $p \in \Pi_L^z$.

        \end{lemma}
        \begin{proof}
            Assume $\mathsf{commit}_M(q,b_m^j)$ holds for $q \in \Pi_{F}^{z}$.
            \textbf{Case 1:} $q$ is honest and the network is synchronous.
            Following Algorithm~\ref{alg_mainchain_protocol}, $q$ generates deterministic sync entries, exchanges them to form $cert_{proc}$ within $\Delta_{sync}$, and broadcasts PROC messages to all $p \in \Pi_L^z$.
            Each $p$ verifies $cert_{proc}$ and updates its local state, so synchronization completes in bounded time.
            \textbf{Case 2:} $q$ is faulty or the network is asynchronous.
            If PROC is not received within $\Delta_{proc}$, a timeout triggers recovery as in Lemma~\ref{lemma:eventual_recovery}.
            The new full member synchronizes its main-chain state, obtains $b_m^j$ and its sync entries, and resumes processing.
            Because faulty members are bounded and asynchrony is finite, the system eventually reaches the conditions of Case 1, at which point all $s_z^i \subset s_m^j$ for $b_{z}^{i}$ where $\mathsf{contains}(b_{m}^{j},b_{z}^i)$ are synchronized to all members $p \in \Pi_{L}^{z}$.

        \end{proof}

        \begin{theorem}[Liveness]
            If a client issues a transaction request to PyloChain, the client eventually receives a corresponding response.

        \end{theorem}

        \begin{proof}
            By Assumption \ref{assumption:correctness_of_underlying_bft}, the client's every request within a zone $z$ submitted to a full member is included in some local block $b_z^i$ and eventually $\mathsf{commit}_L(p, b_z^i)$ holds for $p \in \Pi_L^z$.
            By Lemma \ref{lemma:eventual_main_commit}, this $b_z^i$ is eventually included in a committed main block $b_m^j$.
            By Lemma \ref{lemma:eventual_synchronization}, the corresponding states $s_z^i \subset s_m^j$ are eventually synchronized to all $p \in \Pi_L^z$.
            Each $p$ can then identify the transaction's result and return the response to the client.

        \end{proof}

        \subsubsection{Safety}

        \begin{lemma}[Ordering Consistency]
            \label{lemma:ordering_consistency}

            If $\mathsf{commit}_{L}(p, b_{z}^{i})$ and $\mathsf{commit}_{L}(p, b_{z}^{i'})$ hold for $p \in \Pi_{L}^{z}$ and $i < i'$, then for any main blocks $b^{j}_{m}$ and $b^{j'}_{m}$ such that $\mathsf{contains}(b_{m}^{j}, b_{z}^{i})$ and $\mathsf{contains}(b_{m}^{j'}, b_{z}^{i'})$ hold, it must be that $j < j'$.

        \end{lemma}

        \begin{proof}
            We first note that $b_{z}^{i}$ and $b_{z}^{i'}$ are included in different main blocks, respectively for simplicity.
            $\mathsf{commit}_{L}(p, b_{z}^{i})$ and $\mathsf{commit}_{L}(p, b_{z}^{i'})$ hold for $p \in \Pi_{L}^{z}$ and $i < i'$.
            By Algorithm \ref{alg_mainchain_protocol}, the full member in zone $z$ enforces monotonic ordering during DAG mempool inclusion.
            Specifically, L\ref{alg:AVAIL_start}-\ref{alg:monotonic_ordering_end} ensures $b_{z}^{i}$ must be proposed before $b_{z}^{i'}$ by checking $\mathcal{M}_{z}.$\emph{Next()} = $i$.
            This creates a strict ordering where $b_{z}^{i}$ is included in round $r$ and $b_{z}^{i'}$ in round $r' \ge r$.
            The DAG structure preserves this ordering through its causal dependencies - vertices in round r' reference vertices from round r. When DAG consensus determines the committed sub-DAG via topological sorting, it maintains this round-based ordering.
            Therefore, if $\mathsf{contains}(b_{m}^{j}, b_{z}^{i})$ and $\mathsf{contains}(b_{m}^{j'}, b_{z}^{i'})$ hold, it must be that $j < j'$.
        \end{proof}

        \begin{lemma}[State Synchronization Consistency]

            \label{lemma:state_sync_consistency}
            If $\mathsf{commit}_{M}(q, b_{m}^{j})$ holds for $q \in \Pi_{F}^{z}$ and  $\mathsf{contains}(b_{m}^{j}, b_{z}^{i})$ holds, then  $\mathsf{sync}(p, s^{j}_{m}, s_{z}^{i})$ and $\mathsf{sync}(p', s^{j}_{m}, s_{z}^{i})$ hold for any two honest local members $p, p' \in \Pi_{L}^{z}$.
        \end{lemma}

        \begin{proof}
            Assume that $\mathsf{commit}_{M}(q, b_{m}^{j})$ holds for $q \in \Pi_{F}^{z}$ and  $\mathsf{contains}(b_{m}^{j}, b_{z}^{i})$ holds.
            Since $q \in \Pi_{F}^{z}$ has committed $b_{m}^{j}$, it executes \textsf{ProcessMainBlock}($b_{m}^{j}$) deterministically, generating sync entries $ents_{z}$ for zone $z$. Then $q$ obtains a certificate $cert_{proc}$ with at least $2f_{F}+1$ signatures (including at least $f_{F}+1$ from full members) and broadcasts a PROC message containing $cert_{proc}$ and $ents_{z}$ to all $p \in \Pi_{L}^{z}$.
            All honest local members $p, p' \in \Pi_{L}^{z}$ receive this message, verify $cert_{proc}$, and apply identical $ents_{z}$ to their local states.
            If $q$ is malicious and attempts to send conflicting $ents_{z}$ and $ents'_{z}$ to different local members, the verification of $cert_{proc}$ prevents this attack: the certificate must contain at least $f_{F}+1$ matching signatures from honest full members for the same content, making it impossible to create valid certificates for conflicting entries.
            Therefore, $\mathsf{sync}(p, s^{j}_{m}, s_{z}^{i})$ and $\mathsf{sync}(p', s^{j}_{m}, s_{z}^{i})$ hold for $p \in \Pi_{L}^{z}$.
        \end{proof}

        \begin{corollary}[Global Transaction Consistency]
            \label{corollary:global_tx_consistency}
            Let $tx$ be a global transaction in $b_{m}^{j}$ with write set $W_{tx}$ affecting zones $Z_{A} = \{z_{1}, z_{2}, ..., z_{k}\}$. If $\mathsf{commit}_{M}(q, b_{m}^{j})$ holds for $q \in \Pi_{F}$, then for any two zones $z_{k_1}, z_{k_2} \in Z_{A}$ and any honest local members $p_i \in \Pi_{L}^{z_{k_1}}$ and $p_j \in \Pi_{L}^{z_{k_2}}$, both observe consistent state updates from $tx$.
        \end{corollary}

        \begin{proof}
            Because any $q \in \Pi_{F}$ executes $tx$ deterministically during \textsf{ProcessMainBlock}($b_{m}^{j}$), it produces identical write set. For each affected zone $z_k \in Z_{A}$, \textsf{AppendEntry} generates zone-specific entries $ents_{z_k}$ containing the portion of the write set where its ownership belongs to $z_k$.
            By Lemma \ref{lemma:state_sync_consistency}, all honest $q \in \Pi_{L}^{z_k}$ receive and apply identical $ents_{z_k}$ for each zone independently.
            Since all zones' entries originate from the same deterministic execution of $tx$ and are committed through the same $b_{m}^{j}$, all honest $q \in \Pi_{L}^{z}$ where $z \in Z_{A}$ observe consistent effects of $tx$, ensuring cross-zone transaction consistency.
        \end{proof}

        \begin{theorem}[Safety]
            For any transaction $tx$, if a client observes a finalized result (either success or abort), then no other client will ever see a conflicting result for the same $tx$.
        \end{theorem}

        \begin{proof}
            Let $tx$ be a transaction in $b_{z}^{i}$. By Lemma \ref{lemma:ordering_consistency}, $b_{z}^{i}$ is included in $b_{m}^{j}$ only after $b_{z}^{i'}$ is included in $b_{m}^{j'}$ where $i < i'$ and $j < j'$.
            We then consider two cases based on transaction types:
            Case 1: Assume that $tx$ is local transaction. During main block processing:
            (i) If $tx$ passes validation, it commits with write set $W_{tx}$ if any.
            (ii) If $tx$ is aborted due to interference, $ents_{z}$ for $tx$ would contain the corresponding main states.
            By Lemma \ref{lemma:state_sync_consistency}, all honest $q \in \Pi_{L}^{z}$ apply identical $ents_{z}$, ensuring all clients in zone $z$ observe the same result for $tx$.
            Case 2:
            Assume $tx$ be a global transaction which affects zones $Z_{A}$. By Corollary \ref{corollary:global_tx_consistency}, all honest local members across $Z_{A}$ receive consistent state updates from $tx$'s deterministic execution on the main chain. Therefore, all clients observing $tx$ across $Z_{A}$ see identical state updates.


        \end{proof}

        \subsubsection{Performance Analysis}
        Unlike performance sharding, PyloChain with balanced sharding achieves throughput scalability that increases almost linearly with the number of shards up to a certain point, after which the performance gain becomes limited. It is therefore important to theoretically understand when this saturation point occurs.
        We denote the number of zones by $Z$, the local block generation rate of a full member by $\lambda_{Z}$, the bandwidth limit of a full member by $\gamma$, the maximal local block generation rate under its bandwidth constraint by $\lambda_{Z}^{\gamma}$, the average local block size by $K$, the metadata overhead (e.g., certificate) by $\alpha$, the system throughput with $Z$ zones by $T(Z)$, and the bandwidth consumption of a full member with $Z$ zones by $BC(Z)$. We assume $Z \geq 2$.
        Each full member must receive local blocks from all zones except its own. Hence, its bandwidth consumption is bounded by the maximum available bandwidth $\gamma$, which yields the following relation:
        \begin{equation}
            BC(Z) = (Z-1)\lambda_{Z} \times K \times \alpha \leq \gamma .
        \end{equation}

        The maximum bandwidth consumption can be expressed as:
        \begin{equation*}
            BC_{\max}(Z) = (Z-1)\lambda_{Z}^{\gamma} \times K \times \alpha = \gamma .
        \end{equation*}

        From this relation, we obtain
        \[
            \lambda_{Z}^{\gamma} = \frac{\gamma}{(Z-1) \times K \times \alpha} ,
        \]
        and the system throughput $T(Z)$ grows linearly with the number of zones until the bandwidth becomes saturated, as follows:
        \begin{equation*}
            T(Z) = \min\big(Z\lambda_{Z}, \, Z\lambda_{Z}^{\gamma}\big)
            = \min\left(Z\lambda_{Z}, \, \frac{\gamma}{(Z-1) \times K \times \alpha}\right) .
        \end{equation*}

        We then define the saturation point $Z^{*}$, the number of zones at which performance scaling ceases to be linear, as:
        \begin{align*}
            Z^{*}\lambda_{Z} & = Z^{*}\,\frac{\gamma}{(Z^{*}-1) \times K \times \alpha}, \\
            Z^{*}            & = 1 + \frac{\gamma}{\lambda_{Z} \times K \times \alpha}.
        \end{align*}

        \subsubsection{Storage Cost Analysis}
        We provide a system-wide storage cost model as a function of the number of zones $Z$ under different sharding schemes. For simplicity, we assume that every chain has the same length $B$ and the same average local block size $K$.
        First, in balanced sharding, the storage cost of the main chain maintained by full members is $Z \times |\Pi_{F}^{z}| \times B \times  K $, and the storage cost of local chains maintained by local members is $Z\times |\Pi_{L}^{z}| \times B \times K$. Hence, the total storage cost of balanced sharding is $Z \times (|\Pi_{L}^{z}|+|\Pi_{L}^{z}|) \times B \times K$.
        In performance sharding, only the storage cost of local chains scales with the number of zones, given by $Z \times |\Pi_{L}^{z}| \times (B \times K)$. In availability sharding, every member holds every chain, so the storage cost of the model is $Z \times (|\Pi_{F}^{z}| \times |\Pi_{L}^{z}|) \times (B \times K)$.

        \subsubsection{Auditing Overhead}
        We brifely discuss the auditing mechanism in PyloChain requires no specialized modules beyond lightweight timers and signature verifications for externalized certificates (for availability, DAG inclusion, and main block processing). Its runtime cost is therefore negligible compared to the dominant overhead of all-to-all local block broadcasts and main chain consensus. In our performance model, this cost is naturally subsumed under the metadata factor $\alpha$, since auditing merely adds a small constant-size certificate per main block.

        \section{Evaluation}
        \label{sec:evaluation}
        \subsection{Implementation}
        \label{sec:implementation}
        To evaluate the feasibility of the PyloChain designs, we prototyped PyloChain in GoLang \cite{golang} and integrated it into Hyperledger Fabric \cite{HyperledgerFabricv21} to support smart contracts, incorporating necessary modifications for the PyloChain implementation.
        Local members are implemented as peer nodes, while full members consist of both peer nodes and orderer nodes for block ordering and cross-zone communications. We configured the peer nodes to execute smart contracts locally. We use LevelDB \cite{goleveldb} for state management database.
        For DAG BFT, we utilize the Rust-based implementation of Narwhal/Bullshark \cite{faceboonnarwhal, rustlang} and extend it to support the main chain of PyloChain. Specifically, we configure its number of block distributors (called Workers) to match the number of shards, which directly affects the performance of each full member in concurrently sending and receiving local blocks \cite{narwhaltusk}. We use a partially synchronous version of Bullshark \cite{bullshark} for the consensus logic, rather than its fully asynchronous version, to align with the network assumptions of PyloChain.

        We also implemented other sharding schemes, i.e., availability sharding and performance sharding, within our environment for comparison with PyloChain. For availability sharding, we simply configured full members to propagate the collected local blocks from the DAG mempool to their respective local members within the same zone. For performance sharding, we removed the main chain and implemented a 2PC protocol to handle cross-shard transactions instead. In the 2PC protocol, a designated full member acts as the global coordinator. All cross-shard transactions issued within zones are forwarded to this coordinator, which then carries out the 2PC protocol, i.e., the prepare phase for locking and aggregating the involved states, and the commit phase for finalizing the transaction (i.e., commit or abort) and delivering the finalized results to the involved local chains.


        \subsection{Experimental Setup}
        We conducted our experiments on an in-lab cluster consisting of 24 machines with three different configurations: 6 high-end machines with AMD Ryzen CPU 3990x (2.9GHz) and 256GB RAM, 6 medium-end machines with AMD Ryzen CPU 3970x and 128GB RAM, and 12 low-end machines with AMD Ryzen CPU 5950x (3.4GHz) and 32GB RAM. All machines feature Samsung SSD 970, run Ubuntu 20.04, and are connected via a 10 Gbps Ethernet network.
        We evaluated PyloChain's overall performance in an environment with up to 18 zones, where in each zone, we deployed one full member, four local members, and four clients, which sums up to a total of 18 full members for the main chain, and 72 local members for 18 local chains, respectively.
        All members in our experiment run as Docker containers \cite{dockercontainer}, orchestrated by Docker Swarm.
        We configure the network into multiple overlay networks for the local chains and the main chain over Docker Swarm. Local members participate in a single overlay network within their zone, while full members join two overlay networks, consisting of one for the local chain within a zone and another for the main chain across zones.

        We  set up a micro-payment application using SmallBank \cite{smallbank} to evaluate PyloChain's performance scalability.
        The SmallBank workload is representative of typical blockchain applications, as it primarily consists of state database read/write operations such as \textsf{deposit} (one write) and \textsf{send\_payment} (two reads to check the sender's balance and two writes).
        Note that these I/O patterns on the state database are generic and thus applicable to other common scenarios such as supply chain (e.g., recording inter-party transactions) and healthcare (e.g., logging a sensor value).
        For our setup, 300,000 users were statically initialized and distributed equally among the zones.
        The average size of each transaction and a local block are 2.89KB and 1.4MB, respectively.
        The experimental parameters of the main chain using DAG BFT \cite{faceboonnarwhal} are as follows: each worker batch is set to include one local block. A primary can propose up to 40 digests per vertex. The maximum delay for a primary to propose a vertex is 800ms.
        We measured PyloChain's throughput and latency on the client side, with each client asynchronously submitting transaction requests at a send rate ranging from 10k to 110k, in steps of 10k, evenly distributed across the clients.
        Latency was measured as the time between the client submitting a transaction and receiving a commit event from the local members, while throughput was calculated by summing the number of transactions committed per second as measured by all clients.

        \subsection{Overall Performance}
        \begin{figure*}
            \centering
            \includegraphics[scale=.70]{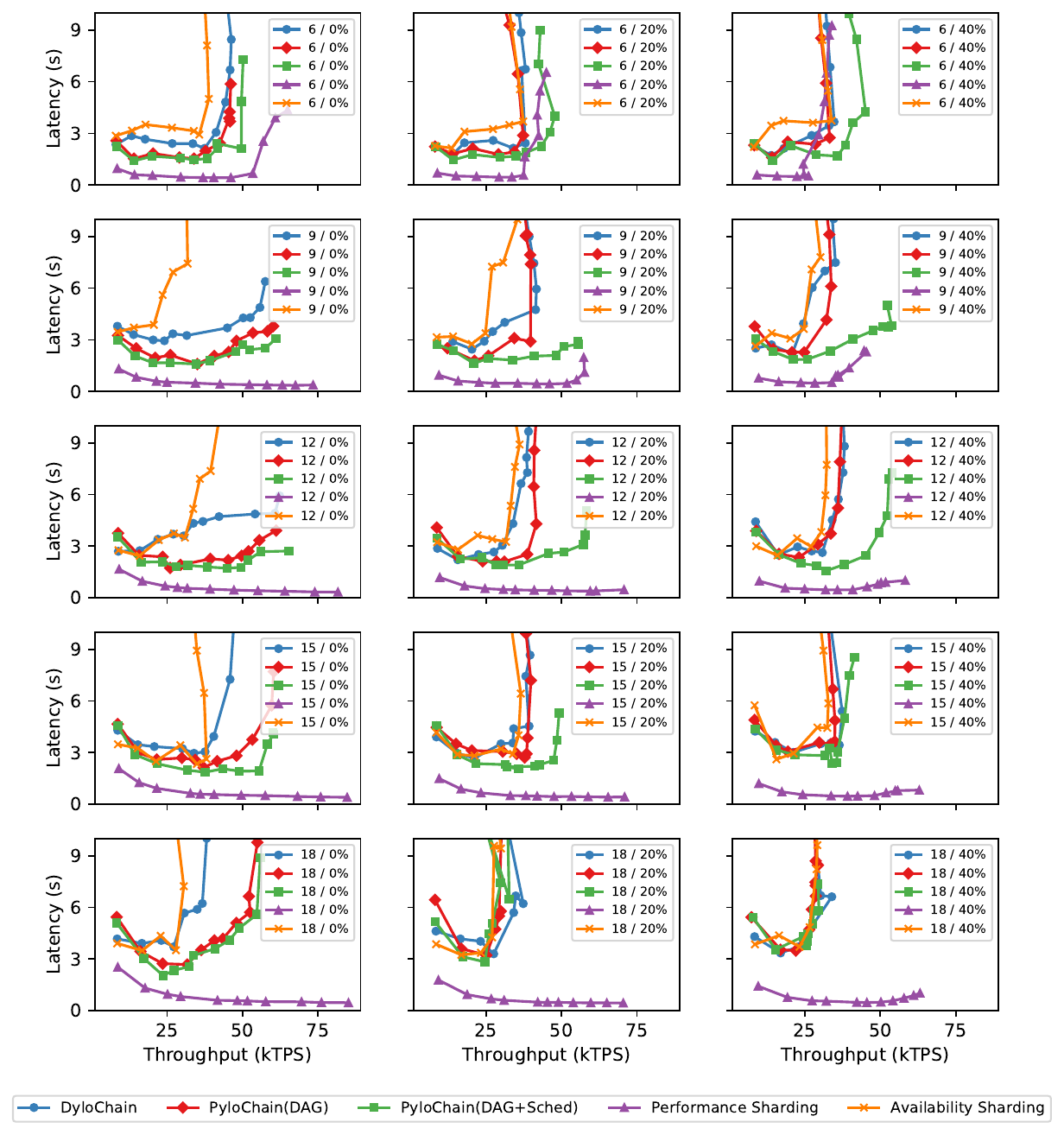} 
            \caption{Overall Performance of PyloChain: Each plot represents the performance in terms of throughput and latency, according to the number of zones and the percentage of global transactions, denoted by \#Zones / \%GlobalTX. For example, "12 / 20\%" indicates that the experiment was conducted in 12 zones with 20\% of the global transactions.} 
            \label{fig:overallperformance} 
        \end{figure*}



        We demonstrate PyloChain's scalability in throughput and latency by varying the number of zones and the global transaction ratio, comparing it with performance and availability sharding (Fig. \ref{fig:overallperformance}). We distinguish two PyloChain versions: PyloChain (DAG), using only DAG-based BFT consensus, and PyloChain (DAG+Sched), adding scheduling technique.
        We first compare PyloChain with DyloChain, then with other sharding schemes.

        \subsubsection{Zone Scalability}

        We first examine all local workloads (0\% global transactions). Notably, PyloChain (DAG) consistently outperforms DyloChain, especially in latency. With 9 zones, PyloChain (DAG)'s latency (2.67s) is 1.47x faster than DyloChain (3.92s), though throughput is similar (PyloChain: 37,432 TPS, DyloChain: 35,148 TPS). With 18 zones, PyloChain (DAG) achieves greater improvements (37,090 TPS, 4.84s latency) compared to DyloChain (30,672 TPS, 8.13s latency).
        This difference arises from main block consensus methods. DyloChain synchronously includes a fixed number of local blocks, increasing latency and reducing throughput as zones scale. PyloChain (DAG)'s asynchronous DAG-based consensus allows flexible inclusion based on varying workloads, improving scalability and efficiency. Performance peaks at 12 zones, then declines beyond 15 zones.

        PyloChain (DAG+Sched) shows minimal performance gains over PyloChain (DAG) with zero global transactions. This is because, despite multi-threaded parallel processing in PyloChain (DAG+Sched), validating already-executed local transactions incurs minimal overhead, thanks to the O-X-O-V model. Note that at higher zones with low send rates, latency increases as clients send fewer transactions, not enough to fill each local block batch size in a timely manner.

        \begin{figure}[t]
            \centering
            \includegraphics[scale=.7]{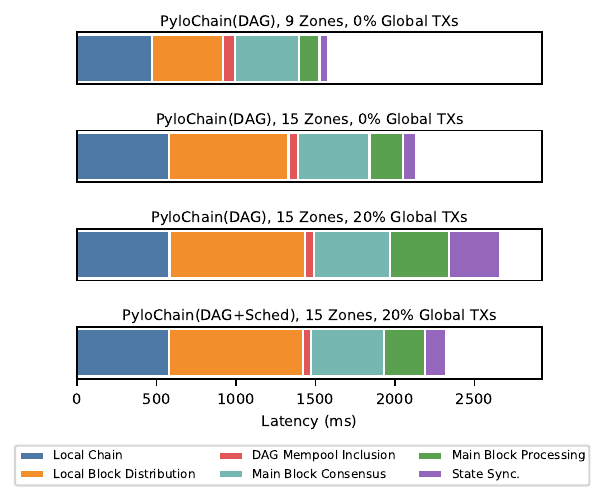} 
            \caption{Latency Analysis from local block confirmation to main chain protocol. Notably, local block distribution cost scales with the number of zones, and global transactions increase main block processing and state synchronization overhead. However, scheduling is observed to mitigate these costs. } 
            \label{fig:latency_analysis} 
        \end{figure}
        \subsubsection{Global Transaction Scalability}
        %

        With 20\% and 40\% global transactions in Fig. \ref{fig:overallperformance}, saturation occurs earlier due to heavier main block processing overhead from global transactions (i.e., smart contract execution, transaction conflicts, and synchronization).
        In these cases, PyloChain (DAG+Sched) consistently outperforms others. For instance, at 12 zones (20\% global transactions), PyloChain (DAG+Sched) achieves 37,642 TPS, 2.80s latency, outperforming PyloChain (DAG) by 1.17x throughput, 1.83x latency, and DyloChain by 1.25x throughput, 1.95x latency.
        At a send rate of 90,000, PyloChain (DAG+Sched) achieves 57,263 TPS with 3.10s latency, offering 1.39x and 1.49x higher throughput, and 2.76x and 2.63x faster speed compared to PyloChain (DAG) and DyloChain, respectively.
        This is because PyloChain (DAG+Sched) improves by scheduling global transactions at the end of each main block, significantly reducing local transaction interferences and synchronization overhead. Nevertheless, when the proportion of global transactions increases, all systems experience performance degradation because the execution of global transactions significantly slows down main block processing.

        \subsubsection{Comparisons with Other Sharding Schemes}

        Comparing PyloChain's balanced sharding with availability and performance sharding, balanced sharding outperforms availability sharding but trails performance sharding. Availability sharding incurs significant overhead due to local block distribution to all members. Performance sharding horizontally scales due to no main chain overhead, even under global transactions.
        PyloChain's scalability is limited by its main chain's all-to-all local block broadcasting, constraining throughput due to higher network bandwidth usage.

        \subsection{Latency Analysis}

        Figure \ref{fig:latency_analysis} presents the phase-by-phase latency of PyloChain at a 60k tx/s send rate, from which we derive several observations.
        First, the latency of local chain consensus remains stable across all scenarios, as it does not depend on the number of zones.
        When the number of zones increases from 9 to 15, local block distribution latency rises significantly because concurrently broadcast local blocks among full members grow quadratically; we observed a 2.02x increase, confirming that local block dissemination is the dominant cost.
        The DAG mempool inclusion phase stays consistently low, as it requires exchanging only lightweight metadata.
        Main block consensus latency grows only slightly, since it is determined every two rounds and depends on the rate at which quorum-sized sets of local blocks enter the DAG.
        Finally, increasing global transactions to 20\% results in a 1.76x increase in main block processing delay and a 4x increase in state synchronization latency, due to sequential execution of global transactions and fetching main states for state-sync construction. With scheduling, the number of aborted local transactions is substantially reduced, leading to a 29.6\% reduction  in main block processing delay and a 59.3\% reduction in state synchronization latency.

        \begin{figure}[t]
            \centering
            \centering
            \includegraphics[scale=.5]{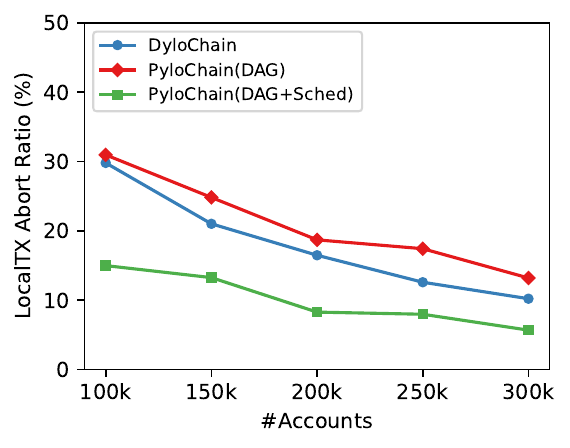}
            \caption{Interference Analysis. PyloChain(DAG) exhibits higher interference than DyloChain, as concurrency degree increases. However, PyloChain(DAG+Sched) with scheduling effectively reduces this interference.}
            \label{fig:abortratio}
        \end{figure}

        \subsection{Interference Analysis}
        We examined transaction interference at 15 zones, 40\% global transactions, and an 80k send rate as in Fig. \ref{fig:abortratio}. PyloChain (DAG) experienced higher abort rates than DyloChain due to larger main blocks. However, PyloChain (DAG+Sched) significantly reduced interference via scheduling. With 300k accounts, PyloChain (DAG+Sched)'s abort ratios (10.20\%, 13.19\%, 5.66\%) were 1.8x and 2.3x lower than DyloChain and PyloChain (DAG).

        \begin{figure}[t]
            \centering
            \subfloat[\scriptsize Identifying Network Bandwidth Bottleneck]{%
                \includegraphics[scale=.5]{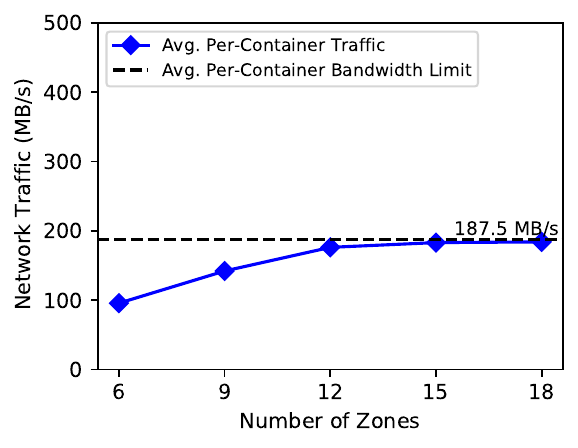}%
                \label{fig:bandwidth_consumption}%
            }\hfill
            \subfloat[\scriptsize Impact of Erasure Coding based Optimization]{%
                \includegraphics[scale=.5]{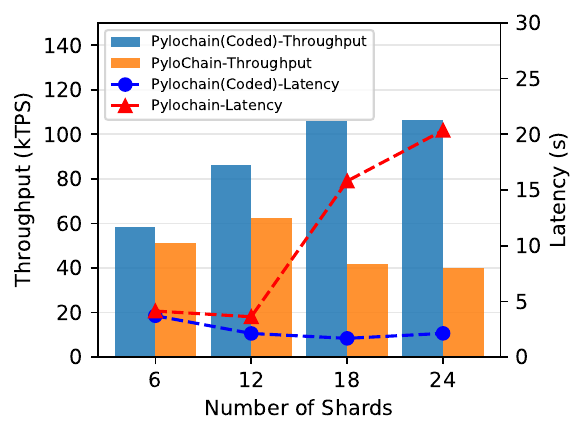}%
                \label{pylochain_fig:coded_blockchain}%
            }
            \caption{Identifying Network Bandwidth Bottleneck and  Impact of Erasure Coding based Optimization. Applying erasure coding to local blocks effectively reduces bandwidth consumption, enabling PyloChain to easily scale up to 24 zones.}
            \label{pylochain_fig:optimization}
        \end{figure}
        \subsection{Optimization}
        We identified the network bandwidth bottleneck shown in Figure \ref{fig:overallperformance} and evaluated the scalability improvements achieved by applying erasure coding-based optimization to PyloChain, which reduces network bandwidth consumption from local block propagation among full members.
        Fig. \ref{fig:bandwidth_consumption} shows the bandwidth consumption as zones scale, confirming that the container bandwidth limits of full members (187.5 MB/s) become saturated at approximately 12 zones, indicating a significant factor hindering zone scalability. Fig. \ref{pylochain_fig:coded_blockchain} demonstrates that applying erasure coding to local blocks effectively reduces bandwidth consumption, enabling PyloChain to easily scale up to 24 zones.
        This experiment reveals that reducing the cost of bandwidth-intensive local block broadcasts introduced by the main chain in hierarchical sharding blockchains has substantial potential impact on performance improvement.

        \begin{figure}[t]
            \centering
            \includegraphics[scale=.5]{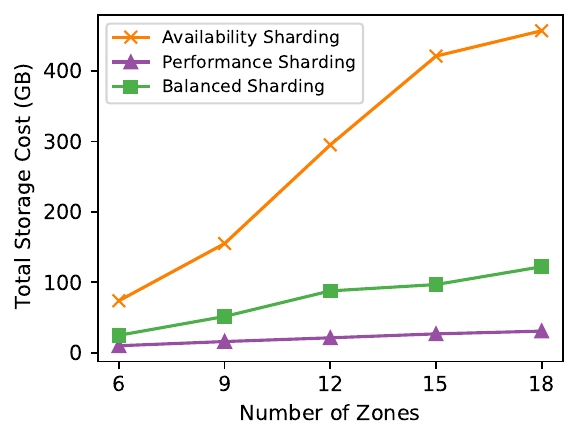}
            \caption{Storage Cost. Availability sharding incurs the highest storage cost as all members store local blocks across zones, while balanced sharding requires only full members to do so, and performance sharding has the lowest cost with each member holding only their respective local chain.}
            \label{fig:pylochain_storage_cost}
        \end{figure}

        \subsection{Storage Cost}
        Fig. \ref{fig:pylochain_storage_cost} compares storage consumption across sharding schemes. In this experiment, clients in each zone submitted transactions at a rate of 4000 per seconds, with all transactions being local, and the experiment lasting one minute.
        As expected, storage costs were highest for availability sharding (up to 457 GB), as local members, along with full members, also store local blocks across zones. Balanced sharding, which requires only full members to store local blocks across zones (up to 122 GB), was closer to the performance of performance sharding, where each member holds only their respective local chain (up to 30.7 GB).


        %


        \begin{figure}[t]
            \centering
            \includegraphics[scale=.7]{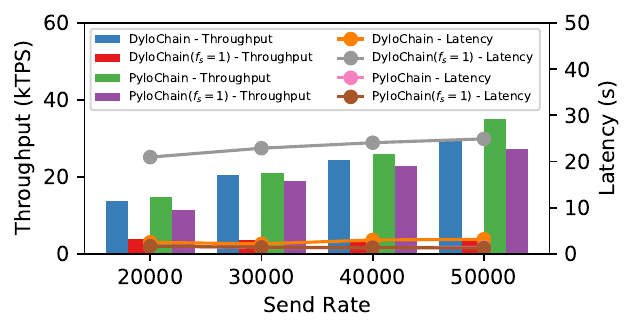} 
            \caption{Slow Zone Resilience. This figure illustrates the performance of PyloChain in a slow zone scenario, highlighting its resilience compared to DyloChain.} 
            \label{fig:slowshardresilience} 
        \end{figure}

        \subsection{Slow Zone Resilience}
        To demonstrate the advantages of PyloChain over a partially synchronous network compared to DyloChain over a synchronous network, we evaluated slow zone resilience. Figure \ref{fig:slowshardresilience} shows results for 9 zones across all workloads. We simulate a slow zone by setting its send rate very low, while varying the rates of the remaining zones from 20k to 50k, and measure throughput and latency.
        The results show that while DyloChain experiences severe degradation when a slow zone exists ($f_z$), PyloChain's performance is affected only by the reduced output of that zone. This is because DyloChain requires one local block from each zone for every main block; when any block is delayed, main blocks are produced only at the fixed timeout (default 1s), thereby causing low throughput and consistently high latency. In contrast, PyloChain uses DAG-based consensus to asynchronously collect local blocks; thus, even if a slow zone lags, the system can form quorums among normal zones and therefore sustain performance with minimal impact.


\section{Discussion}
\label{sec:discussionandfuturework}

\paragraph*{Optimization for Transaction Execution Parallelism}

In PyloChain, global transactions are currently executed sequentially on the main chain, which can become a performance bottleneck as their proportion increases. To alleviate this, PyloChain can adopt deterministic parallel transaction execution techniques \cite{OptMe,BlockSTM} to accelerate global transaction processing. For example, OptME \cite{OptMe}, designed for DAG-based ordering with read/write-set-aware scheduling, can be applied to PyloChain's main chain to identify non-conflicting transaction groups and execute them in parallel. This optimization is limited to global transactions, since the execution order of local transactions is already finalized on their local chains and must be preserved on the main chain. Although a full system integration and evaluation of these algorithms is left for future work, PyloChain's architecture is compatible with such techniques and can exploit global transaction parallelism to improve performance.

\paragraph*{Fully Asynchronous Network and Flexible Consensus}

PyloChain assumes a partially synchronous network model, yet prolonged asynchronous periods or delayed GST can arise in practice, which may limit its applicability in highly variable networks. PyloChain uses a DAG-based BFT protocol for the main chain, where the mempool for data dissemination is fully asynchronous and the consensus layer for total ordering operates under partial synchrony. Tusk \cite{narwhaltusk} is a fully asynchronous consensus protocol that can achieve higher throughput, at the cost of increased latency compared to Bullshark \cite{PartialSyncBullshark,bullshark}, which PyloChain currently employs. Leveraging the modularity of DAG-based BFT, we envision adaptively switching the consensus algorithm according to network conditions, for example using Bullshark under partial synchrony and switching to Tusk during prolonged asynchrony. We leave the design and evaluation of such adaptive consensus selection as future work to enhance the applicability of PyloChain.

\paragraph*{Public Blockchain Applicability}

Although PyloChain is primarily designed for permissioned environments, it can be extended to public blockchains. In permissioned systems, authenticated members naturally support quorum-based BFT consensus, whereas public blockchains must explicitly identify quorum participants to avoid vulnerabilities such as Sybil attacks \cite{SybilAttack}. A natural approach is to use a Delegated Proof-of-Stake (DPoS) based BFT mechanism \cite{SuiStakingUnstaking}, where an elected, staked consensus committee gains voting rights and runs BFT on behalf of the network for a predefined period. PyloChain's hierarchical sharding can then be realized so that the main chain acts as a Layer 1 responsible for data availability and finality, while sharded local chains operate as Layer 2 rollups that perform execution off-chain relative to the main chain.

\section{Conclusion}
\label{sec:conclusion}
In this paper, we propose PyloChain, a hierarchical sharded blockchain that strikes a balance between performance and availability. PyloChain effectively leverages a recent DAG-based mempool to provide local block availability and to achieve efficient main block consensus via locally performed BFT, across sharding zones. It accelerates transaction processing through parallel sharded local chains and a simple scheduling technique to further reduce global transaction interferences. We conduct a systematic evaluation to demonstrate PyloChain's effectiveness.

\bibliographystyle{IEEEtran}  
\bibliography{latex-bib/sample-base}

@online{golang,
  author       = {Google},
  year         = {2025},
  title        = {The Go Programming Language},
  url          = {https://go.dev/},
  month        = Jan,
  lastaccessed = {Mar. 12, 2025}
}

@online{goleveldb,
  author       = {syndtr},
  year         = {2025},
  title        = {LevelDB key\/value database in Go.},
  url          = {https://github.com/syndtr/goleveldb/},
  month        = Jan,
  lastaccessed = {Mar. 12, 2025}
}

@online{rustlang,
  author       = {Rust Team},
  year         = {2025},
  title        = {Rust: A language empowering everyone to build reliable and efficient software.},
  url          = {https://www.rust-lang.org/},
  month        = Jan,
  lastaccessed = {Mar. 12, 2025}
}

@online{smallbank,
  author       = {Bronw Univ.},
  year         = {2019},
  title        = {SmallBank Benchmark},
  url          = {https://hstore.cs.brown.edu/documentation/deployment/benchmarks/smallbank/},
  month        = dec,
  lastaccessed = {Dec. 23, 2019}
}

@inproceedings{HyperledgerFabric,
  author    = {Androulaki, Elli and Barger, Artem and Bortnikov, Vita and Cachin, Christian and Christidis, Konstantinos and De Caro, Angelo and Enyeart, David and Ferris, Christopher and Laventman, Gennady and Manevich, Yacov and Muralidharan, Srinivasan and Murthy, Chet and Nguyen, Binh and Sethi, Manish and Singh, Gari and Smith, Keith and Sorniotti, Alessandro and Stathakopoulou, Chrysoula and Vukoli\'{c}, Marko and Cocco, Sharon Weed and Yellick, Jason},
  title     = {Hyperledger Fabric: A Distributed Operating System for Permissioned Blockchains},
  year      = {2018},
  isbn      = {9781450355841},
  publisher = {Association for Computing Machinery},
  address   = {New York, NY, USA},
  url       = {https://doi.org/10.1145/3190508.3190538},
  doi       = {10.1145/3190508.3190538},
  booktitle = {Proceedings of the Thirteenth EuroSys Conference},
  articleno = {30},
  numpages  = {15},
  location  = {Porto, Portugal},
  series    = {EuroSys '18}
}

@inproceedings{kokoris2018omniledger,
  title        = {Omniledger: A secure, scale-out, decentralized ledger via sharding},
  author       = {Kokoris-Kogias, Eleftherios and Jovanovic, Philipp and Gasser, Linus and Gailly, Nicolas and Syta, Ewa and Ford, Bryan},
  booktitle    = {2018 IEEE Symposium on Security and Privacy (SP)},
  pages        = {583--598},
  year         = {2018},
  organization = {IEEE}
}

@inproceedings{Monoxide,
  author    = {Jiaping Wang and Hao Wang},
  title     = {Monoxide: Scale out Blockchains with Asynchronous Consensus Zones},
  booktitle = {16th {USENIX} Symposium on Networked Systems Design and Implementation ({NSDI} 19)},
  year      = {2019},
  isbn      = {978-1-931971-49-2},
  address   = {Boston, MA},
  pages     = {95--112},
  url       = {https://www.usenix.org/conference/nsdi19/presentation/wang-jiaping},
  publisher = {{USENIX} Association},
  month     = feb
}

@inproceedings{Chainspace,
  author    = {Mustafa Al{-}Bassam and
               Alberto Sonnino and
               Shehar Bano and
               Dave Hrycyszyn and
               George Danezis},
  title     = {Chainspace: {A} Sharded Smart Contracts Platform},
  booktitle = {25th Annual Network and Distributed System Security Symposium, {NDSS}
               2018, San Diego, California, USA, February 18-21, 2018},
  publisher = {The Internet Society},
  year      = {2018},
  url       = {http://wp.internetsociety.org/ndss/wp-content/uploads/sites/25/2018/02/ndss2018\_09-2\_Al-Bassam\_paper.pdf},
  bibsource = {dblp computer science bibliography, https://dblp.org}
}

@inproceedings{castro1999practical,
  title     = {Practical Byzantine fault tolerance},
  author    = {Castro, Miguel and Liskov, Barbara and others},
  booktitle = {OSDI},
  volume    = {99},
  pages     = {173--186},
  year      = {1999}
}

@inproceedings{AHL,
  author    = {Hung Dang and
               Tien Tuan Anh Dinh and
               Dumitrel Loghin and
               Ee{-}Chien Chang and
               Qian Lin and
               Beng Chin Ooi},
  editor    = {Peter A. Boncz and
               Stefan Manegold and
               Anastasia Ailamaki and
               Amol Deshpande and
               Tim Kraska},
  title     = {Towards Scaling Blockchain Systems via Sharding},
  booktitle = {Proceedings of the 2019 International Conference on Management of
               Data, {SIGMOD} Conference 2019, Amsterdam, The Netherlands, June 30
               - July 5, 2019},
  pages     = {123--140},
  publisher = {{ACM}},
  year      = {2019},
  url       = {https://doi.org/10.1145/3299869.3319889},
  doi       = {10.1145/3299869.3319889},
  bibsource = {dblp computer science bibliography, https://dblp.org}
}

@inproceedings{rapidchain,
  author    = {Zamani, Mahdi and Movahedi, Mahnush and Raykova, Mariana},
  title     = {RapidChain: Scaling Blockchain via Full Sharding},
  year      = {2018},
  isbn      = {9781450356930},
  publisher = {Association for Computing Machinery},
  address   = {New York, NY, USA},
  url       = {https://doi.org/10.1145/3243734.3243853},
  doi       = {10.1145/3243734.3243853},
  booktitle = {Proceedings of the 2018 ACM SIGSAC Conference on Computer and Communications Security},
  pages     = {931–948},
  numpages  = {18},
  location  = {Toronto, Canada},
  series    = {CCS '18}
}

@online{HyperledgerFabricv21,
  title    = {Hyperledger Fabric v2.1.0 branch},
  year     = 2020,
  url      = {https://github.com/hyperledger/fabric/tree/v2.1.0
              /},
  addendum = {Accessed: 2022-01-11}
}

@inproceedings{SharPer,
  author    = {Amiri, Mohammad Javad and Agrawal, Divyakant and El Abbadi, Amr},
  title     = {SharPer: Sharding Permissioned Blockchains Over Network Clusters},
  year      = {2021},
  isbn      = {9781450383431},
  publisher = {Association for Computing Machinery},
  address   = {New York, NY, USA},
  url       = {https://doi.org/10.1145/3448016.3452807},
  doi       = {10.1145/3448016.3452807},
  booktitle = {Proceedings of the 2021 International Conference on Management of Data},
  pages     = {76–88},
  numpages  = {13},
  location  = {Virtual Event, China},
  series    = {SIGMOD/PODS '21}
}

@inproceedings{BrokerChain,
  author    = {Huang, Huawei and Peng, Xiaowen and Zhan, Jianzhou and Zhang, Shenyang and Lin, Yue and Zheng, Zibin and Guo, Song},
  booktitle = {IEEE INFOCOM 2022 - IEEE Conference on Computer Communications},
  title     = {BrokerChain: A Cross-Shard Blockchain Protocol for Account/Balance-based State Sharding},
  year      = {2022},
  volume    = {},
  number    = {},
  pages     = {1968-1977},
  doi       = {10.1109/INFOCOM48880.2022.9796859}
}

@inproceedings{bullshark,
  author    = {Spiegelman, Alexander and Giridharan, Neil and Sonnino, Alberto and Kokoris-Kogias, Lefteris},
  title     = {Bullshark: DAG BFT Protocols Made Practical},
  year      = {2022},
  isbn      = {9781450394505},
  publisher = {Association for Computing Machinery},
  address   = {New York, NY, USA},
  url       = {https://doi.org/10.1145/3548606.3559361},
  doi       = {10.1145/3548606.3559361},
  booktitle = {Proceedings of the 2022 ACM SIGSAC Conference on Computer and Communications Security},
  pages     = {2705–2718},
  numpages  = {14},
  location  = {Los Angeles, CA, USA},
  series    = {CCS '22}
}

@inproceedings{Saguaro,
  author    = {Amiri, Mohammad Javad and Lai, Ziliang and Patel, Liana and Loo, Boon Thau and Lo, Eric and Zhou, Wenchao},
  booktitle = {2023 IEEE 39th International Conference on Data Engineering (ICDE)},
  title     = {Saguaro: An Edge Computing-Enabled Hierarchical Permissioned Blockchain},
  year      = {2023},
  volume    = {},
  number    = {},
  pages     = {259-272},
  doi       = {10.1109/ICDE55515.2023.00027}
}

@inproceedings{Ziziphus,
  author    = {Amiri, Mohammad Javad and Shu, Daniel and Maiyya, Sujaya and Agrawal, Divyakant and El Abbadi, Amr},
  booktitle = {2023 IEEE 39th International Conference on Data Engineering (ICDE)},
  title     = {Ziziphus: Scalable Data Management Across Byzantine Edge Servers},
  year      = {2023},
  volume    = {},
  number    = {},
  pages     = {490-502},
  doi       = {10.1109/ICDE55515.2023.00044}
}

@online{faceboonnarwhal,
  title        = {narwhal},
  url          = {https://github.com/facebookresearch/narwhal},
  lastaccessed = {Nov 20, 2023},
  year         = {2021}
}

@article{GriDB,
  author     = {Hong, Zicong and Guo, Song and Zhou, Enyuan and Chen, Wuhui and Huang, Huawei and Zomaya, Albert},
  title      = {GriDB: Scaling Blockchain Database via Sharding and Off-Chain Cross-Shard Mechanism},
  year       = {2023},
  issue_date = {March 2023},
  publisher  = {VLDB Endowment},
  volume     = {16},
  number     = {7},
  issn       = {2150-8097},
  url        = {https://doi.org/10.14778/3587136.3587143},
  doi        = {10.14778/3587136.3587143},
  journal    = {Proc. VLDB Endow.},
  month      = {mar},
  pages      = {1685–1698},
  numpages   = {14}
}

@article{Pyramid,
  author  = {Hong, Zicong and Guo, Song and Li, Peng},
  journal = {IEEE Journal on Selected Areas in Communications},
  title   = {Scaling Blockchain via Layered Sharding},
  year    = {2022},
  volume  = {40},
  number  = {12},
  pages   = {3575-3588},
  doi     = {10.1109/JSAC.2022.3213350}
}

@inproceedings{Meepo,
  author    = {Zheng, Peilin and Xu, Quanqing and Zheng, Zibin and Zhou, Zhiyuan and Yan, Ying and Zhang, Hui},
  booktitle = {2021 IEEE 37th International Conference on Data Engineering (ICDE)},
  title     = {Meepo: Sharded Consortium Blockchain},
  year      = {2021},
  volume    = {},
  number    = {},
  pages     = {1847-1852},
  doi       = {10.1109/ICDE51399.2021.00165}
}

@inproceedings{dyno,
  author    = {Duan, Sisi and Zhang, Haibin},
  booktitle = {2022 IEEE Symposium on Security and Privacy (SP)},
  title     = {Foundations of Dynamic BFT},
  year      = {2022},
  volume    = {},
  number    = {},
  pages     = {1317-1334},
  doi       = {10.1109/SP46214.2022.9833787}
}

@inproceedings{PShard,
  author    = {Gao, Jianbo and Zhang, Jiashuo and Li, Yue and Hao, Jiakun and Wang, Ke and Guan, Zhi and Chen, Zhong},
  title     = {PShard: A Practical Sharding Protocol for Enterprise Blockchain},
  year      = {2023},
  isbn      = {9781450397575},
  publisher = {Association for Computing Machinery},
  address   = {New York, NY, USA},
  url       = {https://doi.org/10.1145/3581971.3581987},
  doi       = {10.1145/3581971.3581987},
  booktitle = {Proceedings of the 2022 5th International Conference on Blockchain Technology and Applications},
  pages     = {110–116},
  numpages  = {7},
  location  = {Xi'an, China},
  series    = {ICBTA '22}
}

@online{Danksharding,
  author       = {Ethereum},
  year         = {2024},
  title        = {Danksharding},
  url          = {https://ethereum.org/en/roadmap/danksharding/},
  month        = Mar,
  lastaccessed = {Mar. 27, 2025}
}

@inproceedings{OHIE,
  author    = {Yu, Haifeng and Nikolić, Ivica and Hou, Ruomu and Saxena, Prateek},
  booktitle = {2020 IEEE Symposium on Security and Privacy (SP)},
  title     = {OHIE: Blockchain Scaling Made Simple},
  year      = {2020},
  volume    = {},
  number    = {},
  pages     = {90-105},
  doi       = {10.1109/SP40000.2020.00008}
}

@online{dockercontainer,
  year         = {2024},
  title        = {Docker},
  url          = {https://www.docker.com/},
  lastaccessed = {Sep. 9, 2024}
}

@article{PeerBFT,
  author   = {Ma, Jeonghyeon and Jo, Yongrae and Park, Chanik},
  journal  = {IEEE Access},
  title    = {PeerBFT: Making Hyperledger Fabric’s Ordering Service Withstand Byzantine Faults},
  year     = {2020},
  volume   = {8},
  number   = {},
  pages    = {217255-217267},
  doi      = {10.1109/ACCESS.2020.3040443}
}

@article{CodedBlockchain24,
  author     = {Yang, Changlin and Chin, Kwan-Wu and Wang, Jiguang and Wang, Xiaodong and Liu, Ying and Zheng, Zibin},
  title      = {Scaling Blockchains with Error Correction Codes: A Survey on Coded Blockchains},
  year       = {2024},
  issue_date = {June 2024},
  publisher  = {Association for Computing Machinery},
  address    = {New York, NY, USA},
  volume     = {56},
  number     = {6},
  issn       = {0360-0300},
  url        = {https://doi.org/10.1145/3637224},
  doi        = {10.1145/3637224},
  journal    = {ACM Comput. Surv.},
  month      = {jan},
  articleno  = {139},
  numpages   = {33}
}

@article{SharDag,
  title   = {SharDAG: Scaling DAG-based Blockchains via Adaptive Sharding},
  author  = {Cheng, Feng and Xiao, Jiang and Liu, Cunyang and Zhang, Shijie and Zhou, Yifan and Li, Bo and Li, Baochun and Jin, Hai},
  journal = {2024 40th IEEE International Conference on Data Engineering (ICDE)},
  year    = {2024},
  address = {Utrecht, Netherlands}
}

@inproceedings{narwhaltusk,
  author    = {Danezis, George and Kokoris-Kogias, Lefteris and Sonnino, Alberto and Spiegelman, Alexander},
  title     = {Narwhal and Tusk: A DAG-Based Mempool and Efficient BFT Consensus},
  year      = {2022},
  isbn      = {9781450391627},
  publisher = {Association for Computing Machinery},
  address   = {New York, NY, USA},
  url       = {https://doi.org/10.1145/3492321.3519594},
  doi       = {10.1145/3492321.3519594},
  booktitle = {Proceedings of the Seventeenth European Conference on Computer Systems},
  pages     = {34–50},
  numpages  = {17},
  location  = {Rennes, France},
  series    = {EuroSys '22}
}

@article{dylochain,
  author    = {Jo, Yongrae and Park, Chanik},
  title     = {A Hierarchical Blockchain supporting Dynamic Locality by Extending Execute-Order-Validate Architecture},
  year      = {2024},
  publisher = {Association for Computing Machinery},
  address   = {New York, NY, USA},
  url       = {https://doi.org/10.1145/3688811},
  doi       = {10.1145/3688811},
  note      = {Just Accepted},
  journal   = {Distrib. Ledger Technol.},
  month     = {aug}
}

@inproceedings{BlockSTM,
  author    = {Gelashvili, Rati and Spiegelman, Alexander and Xiang, Zhuolun and Danezis, George and Li, Zekun and Malkhi, Dahlia and Xia, Yu and Zhou, Runtian},
  title     = {Block-STM: Scaling Blockchain Execution by Turning Ordering Curse to a Performance Blessing},
  year      = {2023},
  isbn      = {9798400700156},
  publisher = {Association for Computing Machinery},
  address   = {New York, NY, USA},
  url       = {https://doi.org/10.1145/3572848.3577524},
  doi       = {10.1145/3572848.3577524},
  booktitle = {Proceedings of the 28th ACM SIGPLAN Annual Symposium on Principles and Practice of Parallel Programming},
  pages     = {232–244},
  numpages  = {13},
  location  = {Montreal, QC, Canada},
  series    = {PPoPP '23}
}

@inproceedings{HieraChain,
  author    = {Tang, Haibo and Zhang, Huan and Zhang, Zhenyu and Zhang, Zhao and Jin, Cheqing and Zhou, Aoying},
  title     = {Towards High-performance Transactions via Hierarchical Blockchain Sharding},
  year      = {2024},
  isbn      = {978-3-031-69576-6},
  publisher = {Springer-Verlag},
  address   = {Berlin, Heidelberg},
  url       = {https://doi.org/10.1007/978-3-031-69577-3_26},
  doi       = {10.1007/978-3-031-69577-3_26},
  booktitle = {Euro-Par 2024: Parallel Processing: 30th European Conference on Parallel and Distributed Processing, Madrid, Spain, August 26–30, 2024, Proceedings, Part I},
  pages     = {373–388},
  numpages  = {16},
  location  = {Madrid, Spain}
}

@misc{PartialSyncBullshark,
  title         = {Bullshark: The Partially Synchronous Version},
  author        = {Alexander Spiegelman and Neil Giridharan and Alberto Sonnino and Lefteris Kokoris-Kogias},
  year          = {2022},
  eprint        = {2209.05633},
  archiveprefix = {arXiv},
  primaryclass  = {cs.DC},
  url           = {https://arxiv.org/abs/2209.05633}
}

@article{MeepoJournal,
  author   = {Zheng, Peilin and Xu, Quanqing and Zheng, Zibin and Zhou, Zhiyuan and Yan, Ying and Zhang, Hui},
  journal  = {IEEE Journal on Selected Areas in Communications},
  title    = {Meepo: Multiple Execution Environments per Organization in Sharded Consortium Blockchain},
  year     = {2022},
  volume   = {40},
  number   = {12},
  pages    = {3562-3574},
  doi      = {10.1109/JSAC.2022.3213326}
}

@inproceedings{OptMe,
  author    = { Ryu, Donghyeon and Park, Chanik },
  booktitle = { 2024 SC24: International Conference for High Performance Computing, Networking, Storage and Analysis SC },
  title     = {{ Toward High-Performance Blockchain System by Blurring the Line between Ordering and Execution }},
  year      = {2024},
  volume    = {},
  issn      = {},
  pages     = {391-406},
  doi       = {10.1109/SC41406.2024.00033},
  url       = {https://doi.ieeecomputersociety.org/10.1109/SC41406.2024.00033},
  publisher = {IEEE Computer Society},
  address   = {Los Alamitos, CA, USA},
  month     = Nov
}

@article{Benzene,
  author   = {Cai, Zhongteng and Liang, Junyuan and Chen, Wuhui and Hong, Zicong and Dai, Hong-Ning and Zhang, Jianting and Zheng, Zibin},
  journal  = {IEEE Transactions on Parallel and Distributed Systems},
  title    = {Benzene: Scaling Blockchain With Cooperation-Based Sharding},
  year     = {2023},
  volume   = {34},
  number   = {2},
  pages    = {639-654},
  doi      = {10.1109/TPDS.2022.3227198}
}

@article{FSBlockchain,
  author   = {Liu, Yizhong and Xing, Xinxin and Cheng, Haosu and Li, Dawei and Guan, Zhenyu and Liu, Jianwei and Wu, Qianhong},
  journal  = {IEEE Transactions on Information Forensics and Security},
  title    = {A Flexible Sharding Blockchain Protocol Based on Cross-Shard Byzantine Fault Tolerance},
  year     = {2023},
  volume   = {18},
  number   = {},
  pages    = {2276-2291},
  doi      = {10.1109/TIFS.2023.3266628}
}

@article{Hiba,
  author    = {Obiri, Isaac Amankona and Gao, Jianbin and Xia, Qi and Xia, Hu and Cobblah, Christian Nii Aflah},
  journal   = { IEEE/ACM Transactions on Networking },
  title     = {{ Hiba: Hierarchical High-Performance Blockchain Architecture }},
  year      = {5555},
  volume    = {},
  number    = {01},
  issn      = {1558-2566},
  pages     = {1-16},
  doi       = {10.1109/TNET.2024.3481488},
  url       = {https://doi.ieeecomputersociety.org/10.1109/TNET.2024.3481488},
  publisher = {IEEE Computer Society},
  address   = {Los Alamitos, CA, USA},
  month     = nov
}

@online{PolygonZkEVMDown,
  author       = {Samyuktha Sriram},
  year         = {2024},
  title        = {Polygon zkEVM Chain Goes Down for 10 Hours.},
  url          = {https://unchainedcrypto.com/polygon-zkevm-chain-goes-down-for-10-hours},
  month        = Mar,
  lastaccessed = {Aug. 26, 2025}
}

@online{SolanaOutages,
  author       = {Lostin},
  year         = {2025},
  title        = {A Complete History of Solana Outages: Causes, Fixes, and Lessons Learnt.},
  url          = {https://www.helius.dev/blog/solana-outages-complete-history},
  month        = Feb ,
  lastaccessed = {Aug. 26, 2025}
}

@online{SuiOutage,
  author       = {Eddie Mitchell},
  year         = {2024},
  title        = {Sui Network Restored Following 2-Hour Outage Ceasing Block Production},
  url          = {https://www.ccn.com/news/crypto/sui-network-outage-halts-block-production},
  month        = Nov,
  lastaccessed = {Aug. 26, 2025}
}

@online{TONOutage,
  author       = {Vince Quill},
  year         = {2025},
  title        = {TON blockchain network back online after brief outage},
  url          = {https://cointelegraph.com/news/ton-back-online-after-outage},
  month        = Jun,
  lastaccessed = {Aug. 26, 2025}
}

@online{AWS_SLA,
  author       = {AWS},
  year         = {2022},
  title        = {Amazon Compute Service Level Agreement},
  url          = {https://aws.amazon.com/compute/sla/?nc1=h_ls},
  month        = May,
  lastaccessed = {Aug. 26, 2025}
}

@online{SuiStakingUnstaking,
  author       = {Sui},
  title        = {Staking and Unstaking},
  url          = {https://docs.sui.io/concepts/tokenomics/staking-unstaking},
  month        = May,
  lastaccessed = {Aug. 26, 2025}
}

@inproceedings{SybilAttack,
  author    = {Douceur, John R.},
  editor    = {Druschel, Peter
               and Kaashoek, Frans
               and Rowstron, Antony},
  title     = {The Sybil Attack},
  booktitle = {Peer-to-Peer Systems},
  year      = {2002},
  publisher = {Springer Berlin Heidelberg},
  address   = {Berlin, Heidelberg},
  pages     = {251--260},
  isbn      = {978-3-540-45748-0}
}

@inproceedings{EclipseAttack,
  author    = {Ethan Heilman and Alison Kendler and Aviv Zohar and Sharon Goldberg},
  title     = {Eclipse Attacks on {Bitcoin{\textquoteright}s} {Peer-to-Peer} Network},
  booktitle = {24th USENIX Security Symposium (USENIX Security 15)},
  year      = {2015},
  isbn      = {978-1-939133-11-3},
  address   = {Washington, D.C.},
  pages     = {129--144},
  url       = {https://www.usenix.org/conference/usenixsecurity15/technical-sessions/presentation/heilman},
  publisher = {USENIX Association},
  month     = aug
}

@inproceedings{Porygon,
  author    = {Chen, Wuhui and Xia, Ding and Cai, Zhongteng and Dai, Hong-Ning and Zhang, Jianting and Hong, Zicong and Liang, Junyuan and Zheng, Zibin},
  booktitle = {2024 IEEE 40th International Conference on Data Engineering (ICDE)},
  title     = {Porygon: Scaling Blockchain via 3D Parallelism},
  year      = {2024},
  volume    = {},
  number    = {},
  pages     = {1944-1957},
  doi       = {10.1109/ICDE60146.2024.00153}
}


\begin{IEEEbiography}[{\includegraphics[width=1in,height=1.25in,clip,keepaspectratio]{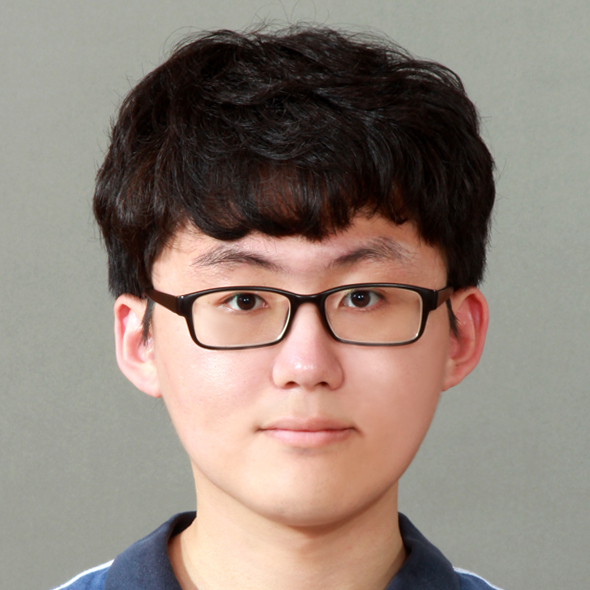}}]{Yongrae Jo} received the B.S. degree in computer science and engineering from Pusan National University, Republic of Korea, in 2017. He is currently pursuing the Ph.D. degree with the Pohang University of Science and Technology (POSTECH). His research interests include distributed systems and blockchain.
\end{IEEEbiography}

\vspace{-33pt}
\begin{IEEEbiography}[{\includegraphics[width=1in,height=1.25in,clip,keepaspectratio]{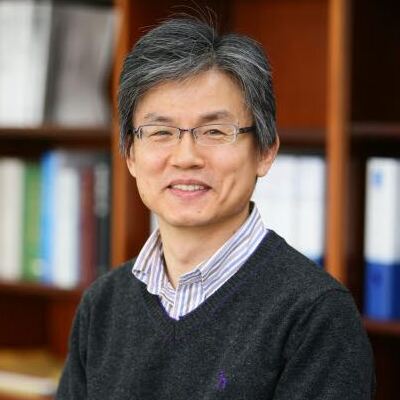}}]{Chanik Park} received the B.S. degree in electronics engineering from Seoul National University, Seoul, Republic of Korea, in 1983, and the M.S. and Ph.D. degrees in electronics and electrical engineering (computer engineering) from KAIST, Daejeon, South Korea, in 1985 and 1988, respectively.
    He was a Visiting Scholar with the Parallel Systems Group, IBM T. J. Watson Research Center, and a Visiting Professor with the Storage Systems Group, IBM Almaden Research Center. He also visited Northwestern University and Yale University, in 2009 and 2015, respectively. Since 1989, he has been working as a Professor with the Department of Computer Science and Engineering, Pohang University of Science and Technology (POSTECH). His research interests include storage systems, operating systems, system security, and blockchain. He has served as a Program Committee Member at a number of international conferences and workshops.
\end{IEEEbiography}


\vspace{11pt}


\vfill

\end{document}